%% file: main.tex
\documentclass[fleqn,10pt]{wlscirep}
\usepackage[utf8]{inputenc}
\usepackage[T1]{fontenc}
\usepackage{amsmath,graphicx}
\usepackage{graphicx} 

\usepackage{booktabs}
\usepackage{longtable}
\usepackage{changes}
\usepackage{makecell} 
\usepackage{footmisc}
\usepackage{xcolor}
\def\keywords{\vspace{.5em}
	{\textit{Keywords}:\,\relax%
}}

\date{May 2024}
\title{The relevance of higher-order ties}

\author[1,*]{Alberto Ceria}
\author[1]{Frank W. Takes}

\affil[1]{Leiden Institute of Advanced Computer Science (LIACS), Leiden University, Einsteinweg 55, 2333 CC Leiden\\

The Netherlands}

\affil[*]{a.ceria@liacs.leidenuniv.nl}

\begin{document}

\begin{abstract}
Higher-order networks effectively represent complex systems with group interactions.
Existing methods usually overlook the relative contribution of group interactions (hyperedges) of different sizes to the overall network structure.
Yet, this has many important applications, especially when the network has meaningful node labels. 
In this work, we propose a comprehensive methodology to precisely measure the contribution of different orders to topological network properties.
First, we propose the order contribution measure, which quantifies the contribution of hyperedges of different orders to the link weights (local scale), number of triangles (mesoscale) and size of the largest connected component (global scale) of the pairwise weighted network.
Second, we propose the measure of order relevance, which gives insights in how hyperedges of different orders contribute to the considered network property. 
Most interestingly, it enables an assessment of whether this contribution is synergistic or redundant with respect to that of hyperedges of other orders. 
Third, to account for labels, we propose a metric of label group balance to assess how hyperedges of different orders connect label-induced groups of nodes. 
We applied these metrics to a large-scale board interlock network and scientific collaboration network, in which node labels correspond to geographical location of the nodes.
Experiments including a comparison with randomized null models reveal how from the global level perspective, we observe synergistic contributions of orders in the board interlock network, whereas in the collaboration network there is more redundancy. 
The findings shed new light on social scientific debates on the role of busy directors in global business networks and the connective effects of large author teams in scientific collaboration networks. 

\end{abstract}

\maketitle


\section{Introduction}
Higher-order networks are an effective way to represent a wide range of complex systems. \cite{battiston2020networks,battiston2022higher,aksoy2020hypernetwork} Differently from traditional pairwise networks, \cite{albert2002statistical,boccaletti2006complex,newman2011structure,newman2018networks} they can explicitly represent relations among consituents that involve more than just pairs.
More generally, group interactions were shown essential to describe a wide range of systems such as human \cite{sekara2016fundamental,cencetti2021temporal,ceria2023temporal} and animal \cite{musciotto2022beyond,iacopini2024not} social networks, collaboration networks, \cite{patania2017shape} drug recombination, \cite{zimmer2016prediction} cellular networks, \cite{klamt2009hypergraphs} species interactions, \cite{levine2017beyond} and the human brain. \cite{petri2014homological, giusti2016two, sizemore2018cliques,santoro2023higher}
Morever, including higher-order interactions evidently affects the collective behaviour of several processes unfolding on the network, such as diffusion, \cite{schaub2020random,carletti2020random} synchronization, \cite{bick2016chaos,skardal2020higher,millan2020explosive,lucas2020multiorder,gambuzza2021stability} contagion, \cite{iacopini2019simplicial,chowdhary2021simplicial,neuhauser2020multibody} and evolutionary\cite{alvarez2021evolutionary,civilini2021evolutionary,civilini2024explosive} processes.

Motivated by these advances, several approaches have been introduced to study the structure of higher-order networks.
Some of these mainly extended the traditional network approaches to include group (higher-order) interactions, such as community detection methods based on generalized modularity,\cite{chodrow2021generative,kaminski2019clustering} spectral clustering \cite{feng2021hypergraph}, bayesian statistics approaches, \cite{contisciani2022inference}  centrality metrics, \cite{benson2019three,feng2021hypergraph,kovalenko2022vector} clustering coefficient, \cite{gallagher2013clustering,klimm2021hypergraphs} and $k$-core decomposition methods. \cite{mancastroppa2023hyper}
On the other hand, new approaches were appositely proposed to characterize group interactions. The overlap of hyperedges with the same or different size was studied, \cite{larock2023encapsulation,landry2024simpliciality} showing that this has influence on the syncronization dynamics unfolding on the network. \cite{malizia2023hyperedge}
Furthermore, the distinction between triads connected by the same hyperedge (closed triangles) and those connected by three different hyperedges (open triangles) allowed previous work to quantify the so called "simplicial closure" phenomenon, i.e., the extent to which triplets of connected nodes are also connected together by a hyperedge of order three or larger.\cite{petri2014homological,benson2018simplicial}

Previous studies on higher-order networks have mostly investigated the impact of group interactions on the overall network structure and on the dynamics of processes unfolding on the network, but rarely zoom in on the separate contribution of different orders.
 As a result, it is currently not possible to investigate how different types of relations (encoded in different hyperedge sizes) contribute to the overall network structure. 
 However, such knowledge is relevant in various application areas. 
 A first example can be found in the study of corporate board interlock networks, \cite{battiston2004statistical,heemskerk2016corporate} in which nodes represent companies and a group interaction (or hyperedge) is a director that connects a set of companies (nodes), because this director is a board member of each of these companies. 
 In these networks, well-studied in the social sciences, it is relevant to investigate how directors with different number of appointments, represented as hyperedges of different orders, contribute to the overall structure of the network, not only in terms of interlocks (pairwise connections), but also if they play a role in connecting/integrating different part of the network in a single connected component.
 This can lead to useful insights, e.g., for a longstanding line of social scientific research on global corporate elites, where it helps understand if networks of interlocking directorates are the result of a corporate elite composed of a small group of "busy directors" (large size hyperedges), or if this is due to a much larger number of interlocking directors with few appointments (small size hyperedges).\cite{stokman1985dutch,fennema2012international,heemskerk2013rise,heemskerk2016global}. 
 A second application can be found in the study of scientific collaboration networks, where authors are connected through hyperedges representing publications by these authors. There, the contribution of links of different orders can be indicative of how smaller or larger author teams contribute to the connectivity and integration of global science. \cite{wu2019large,wuchty2007increasing,lariviere2015team}
 
First methodological steps towards determining the relevance of the ties of different orders were taken by Vasilyeva et al.,\cite{vasilyeva2021multilayer} who proposed a multi-layer network representation to identify the largest size of group interactions that contribute significantly to the network structure, evaluated on several different network properties. A similar objective was also pursued via information-theoretic methods by Lucas et al.\cite{lucas2020multiorder} 
Recently, a filtering procedure to remove small, large or specific orders was proposed by Landry et al.,\cite{landry2024filtering} to investigate how particular global and local network properties are affected when specific orders are preserved or filtered out. With this procedure, it was possible identify nodes that are significantly more central when only specific order hyperedges are included. It was also found that node community membership could be affected by the choice of including or removing hyperedges with specific sizes.

Despite these advances, however, quantifying the contribution made by hyperedges of different orders to the network structure remains an open question. In particular, no metric has been proposed to quantify the contribution of small and large orders on a given topological network property, nor to study how the contribution of small orders is  affected by the contribution of large ones. Furthermore, the precise contribution of different orders to a given topological network measure could vary depending on which measure is considered,\cite{vasilyeva2021multilayer}. This underlines the need of quantifying the contribution of hyperedges with different sizes to network measures that characterize the network structure at different levels of analysis. One can focus on the contribution of different orders to network measures that characterize the structure of the traditional pairwise (or projected) representation of the network, where a link connects each pair of nodes if a hyperedge of any order connects them. This is by far the most adopted network representation in a wide range of applications, including studies on aforementioned board interlock \cite{battiston2004statistical,heemskerk2016corporate,stokman1985dutch,fennema2012international,heemskerk2013rise,heemskerk2016global} and collaboration \cite{newman2001structure,barabasi2002evolution,newman2004coauthorship} networks. At the local level, we are interested in quantifying how hyperedges of different orders contribute to the strength of pairwise connections. At the meso level, it is about how different orders affect clustering, i.e., the number of triangles, and at the macro-level, we may be interested in how hyperedges of particular orders contribute to the emergence of the largest connected component. 
Moreover, in real-world network data, nodes are often assigned meaningful labels, and we may be interested in quantifying the contribution of hyperedges with different orders to the (pairwise and group) connections of nodes with either the same or different labels. 
For example, in the board interlock (scientific collaboration) network such node labels could correspond to the node's geographical location, and these methods could show the different roles played by directors (collaborations) with different number of appointments (author team sizes) in connecting/integrating nodes of a country with those based in the same or in a different country. 

In this work, we propose a new methodology to precisely measure the contribution of orders to the overall network structure, evaluated a. 
First, we present a novel generic \emph{order contribution} measure to quantify the contribution of hyperedges of different orders to a given network property (e.g., link weights, clustering of giant component size). 
Then, we propose a second metric of \emph{order relevance} to quantify the contribution of hyperedges of large and small orders to the different considered network properties.  
Inspired by similar concepts in multivariate information theory\cite{timme2014synergy} and recently also applied in the context of multilayer network analysis\cite{luppi2024quantifying}, the proposed measure allows us to determine whether orders contribute to the considered structural network property in a \emph{synergistic} or in a \emph{redundant} manner. 
Third, to derive meaningful insights from networks in which node labels are available, we propose the measure of \emph{group balance}, which measures how hyperedges of different orders connect nodes with either the same or different labels (i.e., intra- and inter- label connections). Here we focus on the macro level, i.e., how hyperedges of different sizes integrate different sets of nodes with the same label in the largest connected component. 
The fourth and final part of our methodology is an approach to test the significance of the results obtained from the proposed measures, 
in the form of a comparison with two randomized null models that preserve or remove basic properties of the original higher-order network. 

We applied our methodology to two real-world network datasets: a board interlock network of approximately 38M nodes and 13M hyperedges and a scientific collaboration network of approximately 30M nodes and 20M hyperedges. In these networks, each node is labelled according to the country of its headquarter (board interlock network) or research institution (scientific collaboration network). 
We found that in the board interlock network, larger hyperedges, i.e., busy directors, contribute more to the strength of pairwise connections, to the overall number of triangles and to the lergest connected component in the pairwise network than in the collaboration network.
Moreover, our global analysis of the contribution of links of different orders to the largest connected component allows us to distinguish two opposite structural patterns: in board interlock networks, the contribution of large (small) size hyperedges is larger if also small (large) size hyperedges are included. Differently, in the collaboration network, the contribution of large (small) size hyperedges is smaller if also small (large) size hyperedges are included.
Thus, the board interlock and scientific collaboration networks show synergistic and redundant contribution of different orders, respectively.
By comparing these results with two randomized reference models, we observe that the synergistic contribution of hyperedges of different orders observed in board interlock network is highly reduced in models that do not preserve the labels of nodes connected by hyperedges of each order, indicating that the structure of national board interlock networks matters significantly for realizing the global structure, whereas this is to a lesser extent the case in the scientific collaboration network. In this network, instead, the observed value of redundancy cannot be reproduced by models that do not preserve the number of hyperlinks of different orders attached to each node. In the collaboration network, thus, the individual choice of researchers in engaging in collaborations of different sizes in reproducing the redundant contribution of different orders. 
Moreover, the analysis of the order contribution to (group and pairwise) connections of nodes with same or different label/country reveals substantial national differences. In particular in the board interlock network, hyperedges tend to mainly connect companies in the same country when it concerns BRIC countries. 
On the contrary, in countries with clear international orientation such as Luxembourg or Hong Kong, they mostly contribute to connect a few nodes inside the national community with foreign entities, i.e., they create transnational links. 
Finally, in the board interlock, we observe both synergistic and redundant contribution in the way nodes at a country are integrated in the largest connected component of the network. The synergistic contribution is usually also associated with a more relevant contribution of large orders.
In the collaboration network, instead, no synergistic contribution is observed, and a high contribution of large collaborations is usually associated to high redundancy. This means that, in countries where many researchers are connected to the global network by large collaborations, a large portion of these researchers could also be connected by collaborations with fewer members.

In summary, our study proposes methods to investigate the role of interactions with different sizes in higher-order networks, the relation (synergistic or redundant) between contributions of different orders, and metrics to quantify their contribution to connecting nodes with the same or different labels. It enables researchers to explain the detected structural patterns from higher-order data in terms of basic properties of the network by comparing the results with randomized null models.
This provides practitioners with new tools to obtain insights from higher-order network analysis. 

\section{Methodology}
\label{sec:methods}
In this section, we describe our proposed methods.
In Section~\ref{subsec:definitions}, we present basic definitions related to higher-order networks. Then, Section~\ref{subsec:relevance} presents a measure to quantify the contribution of different orders to a given network measure. Three different classical network measures that can be plugged into the proposed measure and capture local, mesoscale, and global structure of the network are discussed in Section~\ref{subsec:traditional}. Then, in Section~\ref{subsec: Community}, we introduce two new metrics to characterize the contribution of hyperedges of different orders to inter- and intra- label connections, from the pairwise and group interaction perspective. Finally, in Section~\ref{subsec:random_ref} we present randomized reference models to investigate if contribution of different orders, quantified according to the measures introduced in Section~\ref{subsec:relevance}, can be reproduced by basic network properties.

\subsection{Higher-order network}
\label{subsec:definitions}
Below, we provide general definitions of higher-order networks using notation also summarized in Table~\ref{tab:merged_metrics} of the Supplementary Material. 
A hyperedge $e$ is a set of nodes $e = \{v_1,\dots v_d\}$, where $d = |e|$ is the size or \emph{order} of the hyperedge.
A static hypergraph \cite{berge1984hypergraphs,bretto2013hypergraph} $H$ is a tuple $(N,E)$, where $N$ is the set of nodes, $E$ is the set of hyperedges. Note that, as a hyperedge is defined as a set of nodes,  $N \supseteq \bigcup_{e\in E} e$.
The traditional pairwise (or projected) representation of the network can be obtained from the higher-order network $H = (N,E)$ as the pairwise network $G = (N,L)$, where any pair of nodes $(v_i,v_j)$ is connected by a link if they are connected by at least a hyperedge of any order in $H$, i.e., $L = \{(v_i,v_j)\in N\times N| v_i,v_j \in e,\ e \in E\}$. We can further assign to each connected pair $(v_i,v_j) \in L$ a weight $w(v_i,v_j) = |\{e\in E| v_i,v_j \in e\}|$, which corresponds to the total number of hyperedges of any order connecting node $v_i$ and node $v_j$.
Moreover, a hypergraph $H' = (N',E')$ is a sub-hypergraph of $H = (N,E)$ if the set $E'$ of its hyperedges is a subset of the set $E$ of the hyperedges of $H$, and its set of nodes $N'$ includes at least all nodes connected by any hyperedge $e \in E'$, i.e., $E \supseteq E'$ and $N\supseteq N' \supseteq \bigcup_{e\in E'} e$.  

In the remainder of the paper, we discuss the contribution of hyperedges of different orders to the network structure. To that end, it is convenient to define the sub-hypergraph composed by hyperedges of a given order only, i.e., from $H = (N,E)$, the sub-hypergraph $H_d$ containing only hyperedges of size $d$, but the entire set $N$ of $H$.  This is the hypergraph $H_d = (N,E_d)$, where $E_d = \{e \in E, |e| = d\}$.
The traditional pairwise (or projected) network obtained from $H_d$, where two nodes are connected by a link if they are connected by at least a hyperedge of order $d$, is $G_d = (N,L_d)$, with $L_d = \{(v_i,v_j)\in N\times N| v_i,v_j \in e,\ e \in E_d\}$.
Finally, the weight $w_d(v_i,v_j) = |\{e\in E_d| v_i,v_j \in e\}|$ is the total number of hyperedges of order $d$ connecting $v_i$ and $v_j$.

\subsection{Order contribution and order relevance of a network measure}
\label{subsec:relevance}

In this section, we first introduce the measure of order contribution to characterize the contribution of each order to a network measure, and then order relevance, a metric that quantifies the relevance of the contribution of large/small order hyperedges to a network measure. 
Relevant notation is summarized in Table~\ref{tab:merged_metrics} of the Supplementary Material. 
Before that, below we give one auxiliary definition and an explanation of which topological network measures the methodology applies to. 

Given a hypergraph $H = (N,E)$, the \textit{$d-$order contribution hypergraph $H^{-}_{d}$} is obtained by including only hyperedges of order $d'\leq d$, i.e., $H^{-}_{d} = (N, \bigcup_{d'\leq d}E_{d'})$.
Hypergraph $H^{-}_{d}$ is equivalent to the one proposed by Vasilyeva et al.\cite{vasilyeva2021multilayer} and obtained by the lower or equal (LEQ) filtering of Landry et al. \cite{landry2024filtering}
Differently, the \textit{inverse $d-$ order contribution hypergraph} $H^{+}_{d}$ is obtained by including only hyperedges with order $d'> d$, i.e., $H^{+}_{d} = (N, \bigcup_{d'> d}E_{d'})$. We indicate a general measure of a higher-order network $H$ with $M(H)$.

The \textit{order contribution} $M^-(d)$ of a network measure $M$ is the value of the quantity $M$ in the $d-$order contribution hypergraph $H^{-}_{d}$, obtained by including only hyperedges of order $d'\leq d$, i.e.,
$M^-(d) = M(H^{-}_{d})$.
Similarly, we define the \textit{inverse order contribution} $M^+(d)$ of the network measure $M$ as the value of the quantity $M$ in the inverse $d-$order contribution hypergraph $H^{+}_{d}$ obtained by including only hyperedges of order $d'> d$, i.e.,
$M^+(d) = M(H^{+}_{d})$.

With the aim of proposing a scalar measure to quantify the impact of small/large orders on a network measure, we consider only network measures for which the order contribution $M^-(d)$ and its inverse $M^+(d)$ are monotonically increasing functions of $d$. \footnote{The methods discussed here can be modified to study the case of monotonically decreasing order contributions, as discussed in detail in the Supplementary Material}. Many network measures satisfy this condition, such as the number of nodes, links, the size of the largest connected component or in general any network measure that is monotonically increasing in the number of links of a network, and is not affected by the presence of disconnected components. Note that, instead, the global clustering coefficient (the ratio between the number of observed triangles and the number of connected triplets of nodes) is not monotonic. 
We then define the \textit{order relevance} $\Gamma_{H}(M)$ with respect to the measure $M$ of the higher-order network $H$ as 
\begin{equation}
    \Gamma_{H}(M) = \frac{\sum_{d = d_{min}} ^ {d_{max}} (M^{-}(d)/M_{max})  - 1}{d_{max} - d_{min}}
\end{equation}
where $M_{max}$ is the maximum value of $M^{-}(d)$, while $d_{max}$ and $d_{min}$ are, respectively, the largest and smallest order of the hyperedges observed in the network.
The order relevance evaluates the impact of different orders by progressively including hyperedges from smallest to larger order.
The values of $\Gamma_{H}(M)$ are bounded between 0 and 1.  As discussed above, the contribution of the order $M^{-}(d)$ is a monotonically increasing function of the order $d$. Thus, when only the smallest order hyperedges $d_{min}$ contribute to the network measure $M$, the order contribution is constant, that is, $M^-(d) = M \ \forall d \in [d_{min},d_{max}]$, which results in a order relevance $\Gamma_{H}(M) = 1$. In contrast, if only the largest order $d_{max}$ contributes to the network measure $M$, then the order contribution is $M^-(d) = M\delta_{d,d_{max}}$, where $\delta_{d,d'}$ is the Kroenecker delta function. In this case, instead, the value of the order relevance is $\Gamma_{H}(M)$ = 0. Thus, values of the order relevance close to 0 correspond to the case in which only large order hyperedges contribute, while values close to 1 correspond to the case in which only small orders contribute.
Note that, in principle, progressively including hyperedges from smaller to larger or larger to smaller size could influence the evaluation of the order relevance.

To investigate the differences in order relevance due to the different strategy of inclusion of hyperedges with different orders, we defined the  \textit{complementary order contribution} $\overline{M^-(d)} = M_{max} - M^+(d)$.
The complementary order contribution $\overline{M(d)}$ of the measure $M$ accounts for the contribution of orders $d'\leq d$, when we progressively include orders from the largest to the smallest.

The \textit{complementary order relevance} is then
\begin{equation}
    \overline{\Gamma_{H}(M)} = \frac{\sum_{d = d_{min}} ^ {d_{max}} (\overline{M^-(d)}/M_{max})  - 1}{d_{max} - d_{min}}.
\end{equation}
In general, the values of the order relevance and the complementary order relevance are different. If $\Gamma_{H}(M) \neq \overline{\Gamma_{H}(M)}$, then the contribution of small (large) orders to the network metric $M$ changes if large (small) orders are also included. To quantify such difference, we introduce the \emph{order relevance gap} as 
\begin{equation}
    \Delta_{H}(M) = \Gamma_{H}(M) - \overline{\Gamma_{H}(M)}.
\end{equation}
The value of $\Delta_{H}(M)$ can distinguish among three different scenarios.
If $\Delta_{H}(M) = 0$, then $\Gamma_{H}(M) = \overline{\Gamma_{H}(M)}$: the contribution of each order to the network metric $M$ is independent from how hyperedges of different orders are progressively included.
Differently, if $\Delta_{H}(M) < 0$, or equivalently $\overline{\Gamma_{H}(M)} > \Gamma_{H}(M)$, then the contribution of hyperedges of large orders to the network metric $M$ is larger if smaller orders are also included. Equivalently, the contribution of small order hyperlinks is also larger if large orders are included. Orders thus contribute \emph{synergistically} to measure $M$. Oppositely, if $\Delta_{H}(M) > 0$, then the contribution of larger (smaller) hyperedges is reduced if smaller (larger) orders are included: the different orders contribute thus \emph{redundantly} to the network measure $M$.

When $\Delta_{H}(M)\neq 0$, the values of the order contribution and of the complementary order contribution are different, and so are the corresponding order relevances.
This means that neither the order contribution $M^-(d)$ and relevance $\Gamma_{H}(M)$ nor the corresponding complementary values $\overline{M^-(d)}$ and $\overline{\Gamma_{H}(M)}$ can unequivocally quantify the contribution and the relevance of hyperedges of different orders to the considered network metrics.
We thus finally define the \emph{average order relevance} $\langle \Gamma_{H}(M) \rangle$ as the average of the order relevance and its corresponding complementary value, i.e.,
\begin{equation}
    \langle \Gamma_{H}(M) \rangle = \frac{\Gamma_{H}(M) + \overline{\Gamma_{H}(M)}}{2} = \frac{\sum_{d=d_{min}} ^ {d_{max}} ( \langle M^-(d) \rangle/M_{max}) - 1}{d_{max} - d_{min}}
\end{equation}
where $\langle M^-(d) \rangle = \frac{M^-(d) + \overline{M^-(d)}}{2}$ is the \emph{average order contribution}.
Given a higher-order network and a monotonic network measure $M$, the average order contribution unequivocally quantifies the separate contribution of each order to the network measure, the corresponding order relevance quantifies the relevance of the contribution of small and large orders, while the order gap measures the synergic or redundant contribution of small and large order hyperedges.

\subsection{Traditional network measures}
\label{subsec:traditional}
In this section we introduce the main network measures for which order relevance gives a meaningful measurement of the contribution of different orders to the value of these measures. Such measures describe the network structure at different levels: link weights at the local (Section~\ref{subsubsec:link_weights_def}), the number of triangles at the meso (Section~\ref{subsubsec:triangles_def}), and the size of the largest connected component (Section~\ref{subsubsec:component_def}) at the global level. Notation related to these measures is summarized in Table~\ref{tab:merged_metrics} of the Supplementary Material. 
\subsubsection{Link weights}
\label{subsubsec:link_weights_def}
As networks are typically studied by adopting a (weighted) pairwise representation, we start by investigating how hyperlinks of different orders contribute to the weights of the projected pairwise topology of a higher-order network.
The first network measure we consider is thus the sum $\Lambda$ of the link weights in the projected network.
Given a higher-order network $H = (N,E)$, the sum $\lambda (d)$ of link weights in the projected graph due to hyperedges of size $d$ only as $\lambda (d) = \frac{1}{2}\sum_{v_i,v_j\in N} w_d(v_i,v_j)$.
We can then define the order contribution of hyperedges of size smaller or equal to $d$ to the sum of link weights of the projected network as:
\begin{equation}
 \Lambda ^{-}(d) = \sum _ {d'\leq d} \lambda(d')    
\end{equation}

The order contribution $\Lambda^-(d)$ is the order contribution $M^-(d)$ introduced in Section~\ref{subsec:relevance} when $M = \Lambda$.
This is clearly a monotonic function of $d$, with values $0\leq \Lambda ^{-}(d) \leq \Lambda$. 
By substituting the general network metric $M$ with $\Lambda$, we can also compute the complementary order contribution $\overline{\Lambda^-(d)}$.
Then, the impact of large/small order hyperedge on the sum of the link weights $\Lambda$ can be computed by the order relevance $\Gamma_{H}(\Lambda)$ and the corresponding complementary measure $\overline{\Gamma_{H}(\Lambda)}$. 
Note that in this case, the value of the order relevance is equivalent to its complementary by definition, i.e.,  $\Gamma_{H}(\Lambda) = \overline{\Gamma_{H}({\Lambda})}$. This is because the sum of link weights obtained by including hyperedges from the smallest to the largest order is equivalent to including them from the largest to the smallest order. The contribution of each hyperedge size is thus not influenced by the presence (or absence) of hyperedges of larger/smaller size.

\subsubsection{Number of triangles} 
\label{subsubsec:triangles_def}

To investigate the contribution of hyperedges to the mesoscale structure of the networks, we compute how they contribute to the number $\tau$ of triangles formed in the projected graph.
The number of triangles impacts the classical measures of clustering (average local and global clustering coefficient), but, differently from these measures, it is a monotonic function in the number of links of a network, and, consequently, in the size $d$ of included hyperedges. This allows us to compute the order contribution and relevance measures described in Section~\ref{subsec:relevance}.

Given a higher-order network $H = (N,E)$, the corresponding $d-$order hypergraph $H^{-}_{d}$ and its inverse $H^{+}_{d}$, we define the corresponding projected graphs $G^{-}_{d} = (N, L^{-}(d))$ and $G^{+}_{d} = (N, L^{+}(d))$ with $L^{-}(d) = \bigcup_{d'\leq d} L_{d'}$ and  $L^{+}(d) = \bigcup_{d'>d} L_{d'}$, respectively. 

Then, we denote the number of triangles in the projected graph $G$ of $H$ as $\tau (G) = \tau(H)$.
The order contribution $\tau^-(d)$  to the number of triangles is then the number of triangles in the projected graph $G^{-}_{d}$ of the $d-$ order contribution hypergraph $H^{-}_{d}$,  constructed by including all hyperedges of order $d'\leq d$.
The inverse order contribution $\tau^+(d)$ is instead the number of triangles in the projected graph $G^{+}_{d}$ of the hypergraph $H^{+}_{d}$ which includes only hyperedges of order $d'>d$.
From the order contribution $\tau^-(d)$ and the inverse $\tau^+(d)$, the complementary order contribution $\overline{\tau^-(d)}$, the corresponding order relevance $\Gamma_{H}(\tau)$ and its complementary $\overline{\Gamma_{H}(\tau)}$ can be obtained following Section~\ref{subsec:relevance}, and substituting the general network metric $M$ with $\tau$.

\subsubsection{Largest connected component}
\label{subsec:def_component}
\label{subsubsec:component_def}

The possible contribution of hyperedges of different orders to the overall network structure also plays at the macro level.
For example, hyperedges of a given size could be more likely to bridge areas of the networks that otherwise would be disconnected from each other.

We quantify the role that hyperedges of different orders have in integrating the network, by computing the order contribution and its corresponding inverse of the largest connected component of the corresponding projected network. 
To quantify the contribution of different orders to the largest connected component of the projected network, we define the order contribution $\sigma^-(d)$, which is the size of the largest connected component of the pairwise projected network $G^-_d$ obtained from the $d-$order contribution hypergraph $H^{-}_d$ which includes only hyperedges of size $d'\leq d$. Equivalently, the inverse order contribution of the largest connected component $\sigma^+(d)$ is the size of the largest connected component of the pairwise projected network $G^-_d$ obtained from the inverse $d-$order contribution hypergraph $H^{+}_d$, where only hyperedges of order $d'>d$ are included.
From the order contributions $\sigma^-(d)$ and its inverse $\sigma^+(d)$, we then follow the definitions in Section~\ref{subsec:relevance} and compute the complementary order contribution $\overline{\sigma(d)}$, the order relevance $\Gamma_{H}(\sigma)$ and its complementary $\overline{\Gamma_{H}(\sigma)}$, and the order relevance gap $\Delta (\sigma)$.
Note that, in this case, the order relevance gap $\Delta_{H}(\sigma) \neq 0$. Consequently, the contribution of small and large orders to $\sigma$ is quantified through the average order contribution $\langle \Gamma_{H}(\sigma)\rangle$. The analysis of the average order relevance and the order gap is further discussed in Section $\ref{subsubsec:connected_overall}$.

\subsection{Inter-/intra-label connection metrics}
\label{subsec: Community}

So far, we have discussed general methods to investigate how hyperedges of different orders contribute to the considered measures of the overall network structure.
In real data, however, additional metadata beyond node connections could be available, e.g., node labels. In this paper, we focus on the particular case where each node is assigned a single label, so that the labels constitute a partition of the network.
As nodes in the hypergraph and the projected network introduced in the previous section are labelled, we are also interested in investigating how hyperedges with different orders tend to connect nodes belonging to the same or different groups induced by these labels. In the remainder of the paper, we use the term \textit{intra-label} to refer to the connections among nodes with the same label, while we use \textit{ inter-label} to indicate the connections among nodes with different labels.

\subsubsection{Intra-/inter-community link weights}
For each node label $\ell$, we introduce the order contribution to the intra-label link weights $\lambda_{\ell,intra}(d)$ as the sum of the weights of links connecting two nodes with the same label $\ell$. Equivalently, we define the inter-label link weights ($\lambda_{\ell,inter}(d)$), which is the sum of the weights of the links connecting nodes with the label $\ell$ with nodes with different labels. 
For example, let us assume a hyperedge of order 9 connecting 8 nodes with label $\ell_1$ and one node with label $\ell_2$. The hyperedge does not contribute to the intra-label link weights of label $\ell_2$, contributes as $\binom{8}{2}$ to the weights of the intra-label links of label $\ell_1$ and as 8 to the weights of inter-label links between label $\ell_1$ and $\ell_2$. 
From  $\lambda_{\ell,intra}(d)$ and $\lambda_{\ell,inter}(d)$ we can compute the corresponding order contribution for the intra- and inter-label link weights as
\begin{equation}
    \Lambda^{-}_{\ell,intra}(d) = \sum_{d'\leq d}\lambda_{\ell,intra}(d')
\end{equation}
and 
\begin{equation}
    \Lambda^{-}_{\ell,inter}(d) = \sum_{d'\leq d}\lambda_{\ell,inter}(d').
\end{equation}
Finally, from the order contributions $\Lambda^{-}_{\ell,intra}(d)$ and $\Lambda^{-}_{\ell,inter}(d)$, we can obtain the corresponding order relevance $\Gamma_{H}(\Lambda_{\ell,intra})$ and $\Gamma_{H}(\Lambda_{\ell,inter})$ as prescribed in Section~\ref{subsec:relevance}.
Also in this case, as previously discussed for the order relevance of the link weights of the overall network $\Lambda$ (see Section ~\ref{subsubsec:link_weights_def}), the order relevance of intra- and inter-label link weights are not influenced by the strategy of inclusion of hyperedges (from small to large sizes or from large to small), so that both order relevance $\Gamma_{H}(\Lambda_{\ell,intra})$ and $\Gamma_{H}(\Lambda_{\ell,inter})$ are equal to their corresponding complements $\overline{\Gamma_{H}(\Lambda_{\ell,intra})}$ and $\overline{\Gamma_{H}(\Lambda_{\ell,inter})}$.

\subsubsection{Intra-/inter- label hyperedges}
\label{subsec:label_group_def}

Beside the order relevance, we can study how hyperedges of different orders connect the nodes with a given label to the nodes with another label by defining, for each label $\ell$, the \textit{label group composition probability} $P_{\ell}(d,k)$ that a random hyperedge connect $d$ nodes, whose $k$ have the same label $\ell$.
Note that, since the number $k$ of nodes with the same label connected by a hyperedge cannot be larger than the size $d$ of the hyperedge itself, $P_{\ell}(d,k) = 0$ as long as $k>d$. 
The average number of nodes with label $\ell$ connected by a hyperedge of order $d$ is then the conditional average $\langle k_{\ell} \rangle_d = \sum_{k}k\ P_{\ell}(k|d)$, where $ P_{\ell}(k|d) = P_{\ell}(d,k)/P_{\ell}(d)$ and 
$P_{\ell}(d) = \sum_kP_{\ell}(d,k)$. Note that $P_{\ell}(d)$ is the probability that a random set of any number of nodes with label $\ell$ is connected by a hyperedge of order $d$. 
Given the label group composition probability $P_{\ell}(d,k)$, the two extreme cases $P_{\ell}(d,k) = \delta_{k,d}P_{\ell}(d)$  and $P_{\ell}(d,k) = \delta_{k,1}P_{\ell}(d)$ correspond to the cases where the hyperedges always connect nodes with label $\ell$ to nodes with same or different label, respectively. To allow an easy comparison of the group composition distributions of different labels, we define the \textit{group balance} of a label $\ell$
\begin{equation}
    \Phi_{\ell} = \frac{\sum_{d}\sum_kP_{\ell}(d,k)\  (d-k)}{\sum_d\sum_kP_{\ell}(d,k)\ (d-1)}
\end{equation} which measures the difference of the label group composition probability $P_{\ell}(d,k)$ from the case in which $P_{\ell}(d,k) = \delta_{d,k}P_{\ell}(d)$.
 This measure has value $0\leq \Phi_{\ell} \leq 1$, where $\Phi_{\ell} = 0 $ if $P_{\ell}(d,k) = \delta_{k,d}P_{\ell}(d)$ and 
$\Phi_{\ell} = 1 $ if $P_{\ell}(d,k) = \delta_{1,k}P_{\ell}(d)$.
Finally, it should be noted that as the label group balance approximates with a scalar number the tendency of hyperedges of different size to interact in groups with nodes with same or different labels, it can be strongly affected by the balance of low orders connections (as they are more frequent) and may fail to capture the contemporary tendency of nodes to interact more frequently as a minority or a majority in groups of different sizes.\cite{veldt2023combinatorial} The effects of these limitations on the label group balance can be taken into account by also studying the full label group composition probability.
\subsubsection{Labelled nodes in the largest connected components}
\label{subsec:label_component_def}

In Section ~\ref{subsubsec:connected_overall} we present an analysis of the largest connected component, together with the corresponding definitions of order relevance, its complementary, and the order relevance gap.
However, besides the overall contribution of hyperedges of different orders, we can further ask how hyperedges of different orders integrate the different groups of nodes with the same label in the overall largest connected component.
We thus defined the number of nodes with label $\ell$ in the largest connected component of the projected network as $\sigma_{\ell}$, and the corresponding order contribution $\sigma^-_{\ell}(d)$, indicating the number of nodes with label $\ell$ in the largest connected component of the projected network of the $d-$order contribution hypergraph $H^{-}(d)$, obtained by including only hyperedges of size $d'\leq d$. 
In principle, the order contribution $\sigma^-_{\ell}(d)$ and its corresponding inverse  $\sigma^+_{\ell}(d)$ are not monotonic in $d$.
We denote the set of nodes in the largest connected components of $H^{-}(d)$ and $H^{+}(d)$ as $N_{\sigma}(H^{-}(d))$ and $N_{\sigma}(H^{-}(d))$. The order contribution $\sigma^-_{\ell}(d)$ is however a monotonic function of $d$ as long as $N_{\sigma}(H^{-}(d)) \subseteq N_{\sigma}(H^{-}(d'))$ for each couple $d'\leq d$. Equivalently, the inverse order contribution $\sigma^+_{\ell}(d)$ and $\overline{\sigma^-_{\ell}(d)}$ are monotonic as long as $N_{\sigma}(H^{+}(d)) \subseteq N_{\sigma}(H^{+}(d'))$ for each couple $d'\geq d$. We verified that this condition is approximately satisfied for each considered higher-order network.\footnote{The condition is usually satisfied, with the few exceptions in the board interlock network. This produces minimal effects on the results. For further details, see the dedicated Section in the Supplementary Material.}
From $\sigma^-_{\ell}(d)$, we follow the definitions of Section ~\ref{subsec:def_component} and introduce the inverse order contribution $\sigma^+_{\ell}(d)$, the complementary $\overline{\sigma^-_{\ell}(d)}$, the order relevance $\Gamma_{H}(\sigma_{\ell})$, the complementary order relevance $\overline{\Gamma_{H}(\sigma_{\ell})}$ and the order relevance gap $\Delta_{H}(\sigma_{\ell})$.

\subsection{Network randomized reference models}  
\label{subsec:random_ref}

To determine the significance of the values of the order contribution and/or relevance of the considered metrics obtained from real data, we compare such results with randomized reference models that can preserve or destroy specific properties of the network. 
In particular, we use models that can sample random hypergraphs while preserving basic properties of the original network. The two proposed randomized null models obtained from a hypergraph $H$ are  $H^1$,  $H^2$.
Both randomized null models preserve approximately the total number of hyperlinks attached to each node (its degree in the higher-order network) and the size of each hyperlink. For completeness's sake, we note that if, as a result of the randomization, a hyperedge of order $d$ contains duplicate nodes, then the two nodes are merged and the hyperedge turns into an order $d-1$ hyperedge. 

The first one, i.e., $H^1$, is obtained by repetitively swapping the nodes between random pairs of hyperedges with the same order.
In this way, the number of hyperedges of each order attached to each node and consequently the number of hyperedges of each size present in the network are preserved.
However, in this way, the labels of nodes connected by any hyperedge are not preserved, thus the label group composition probability $P_{\ell}(d,k)$ is not preserved by this randomization. The randomized null model $H^1$ is then used to investigate if the number of hyperedges of each order attached to each node,  without label composition of hyperedges described by $P_{\ell}(d,k)$, can reproduce the results obtained from real data. 

To preserve instead the group label composition probability $P_{\ell}(d,k)$,  we introduce the model $H^2$, which corresponds to a configuration model for labelled hypergraphs. \cite{chodrow2020annotated}
In this case, for each pair of hyperedges,  we randomly swap one of their nodes with another one with the same label.
This ensures that the number of nodes of each label in each hyperedge is preserved, thus preserving the label composition probability $P_{\ell}(d,k)$ . In this randomization, however the number of hyperedges of each order attached to each node is not preserved.

\section{Data} 
\label{sec:data}
We applied our methods to investigate the contribution of different orders on different measures that characterize the network structure of two real-world networks: a board interlock network and a collaboration network.
\subsection{Board interlock network}
The data were collected from the Orbis Bureau van Dijk dataset \footnote{orbis.bvdinfo.com} in December 2017. Orbis Bureau van Dijk is a comprehensive resource containing corporate information globally, gathered from official country registries. For an extensive investigation on the overall quality and limitations of the data source, we refer to the work of Garcia-Bernardo and Takes \cite{garcia2018effects}, and Heemskerk et al. \cite{heemskerk2018promise} 
Generally, the data of large economies and major corporations are of higher quality, with the exception of North American corporate boards, for which data quality is moderate. From the company appointment register, we selected only individuals holding current positions as executive boards, supervisory boards, and boards of directors, following the approach of Valeeva \cite{valeeva2022backbone}.
We considered only directors appointed by a maximum of 50 different companies.
As we are interested in board interlocks, individuals appointed by a single company are discarded. 

We build the higher-order board interlock network where each node is a company and each individual director appointed by a set of companies is represented as a hyperedge connecting the corresponding nodes. We disregard cases in which multiple directors (hyperedges) interlock the exact same set of nodes/companies, that is, the topology of the higher-order board interlock network is unweighted.
As each company is assigned a unique country label, geographical location labels induce a partition of the nodes.
The topological properties of the board interlock network are presented in Table~\ref{tab:details}, while the distribution of the number of hyperedges of each size and the number of nodes with $k$ hyperedges attached is presented in Figure ~\ref{fig:overall_stat}.

\subsection{Scientific Collaboration Network}
\begin{figure}[ht]
    \centering
    \includegraphics[width=0.8\linewidth]{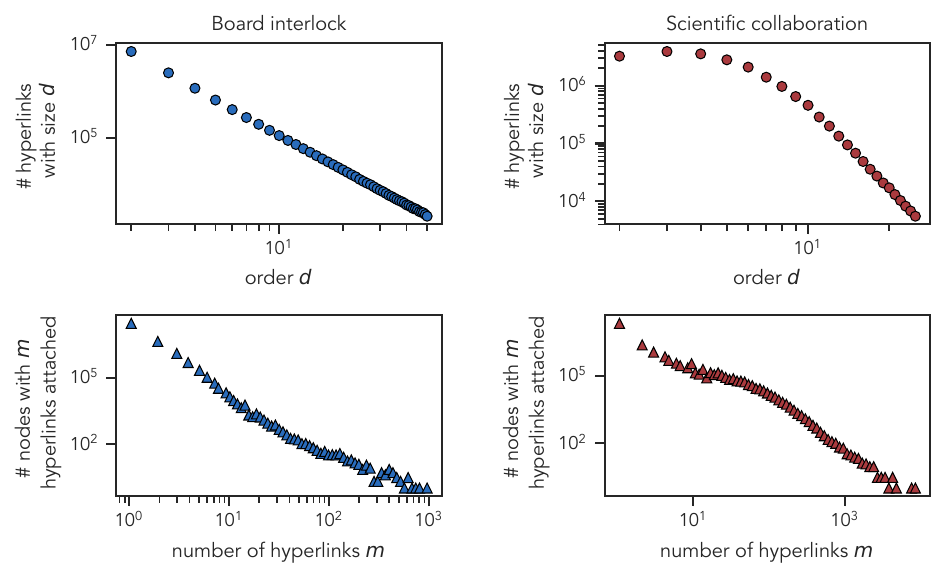}
    \caption{\textbf{Network size}. The number of hyperedges of size $d$ (dots) and the number of nodes with $m$  hyperedges  attached to it (triangles) for board interlock (blue) and collaboration network (red). In the bottom two plots, the horizontal axis is divided in 80 bins.}
    \label{fig:overall_stat}
\end{figure}

The second considered dataset is based on the Clarivate’s Web of Science database (WoS). In particular, this version was collected from the Centre for Science and Technology
Studies (CWTS) at Leiden University in 2023. The original WoS dataset has been enriched by CWTS in different ways, such as through their own consistent and precise assignment of publications to universities and organizations, \cite{waltman2012leiden} geocoding
of the author's addresses, and improved author disambiguation.\cite{caron2014large}
We consider publications released in 2008–2023 categorized as Article,
Review, Letter or Proceeding Paper. Publications with missing author-affiliation
linkages or missing both geo-location and organization information are excluded.
We then selected only papers with a number of authors equal or smaller than 25.
Moreover, for each node/author, we assigned the label corresponding to the country location of the research affiliation associated with the largest number of papers published by the author.
The topological properties of the scientific collaboration network are presented in Table~\ref{tab:details}, while the distribution of the number of hyperedges of each size and the number of nodes with $k$ hyperedges attached is presented in Figure~\ref{fig:overall_stat}.
{\setlength\tabcolsep{4.5pt}
\begin{table}[ht]
\centering
\begin{tabular}{lrrrrrrr}
\toprule
\textbf{Network} & \textbf{$|N|$} & \textbf{$|L|$} & \textbf{$|E|$} & \textbf{$\tau$} & \textbf{$\sigma$} & \textbf{$\Lambda$} & \# labels \\
\midrule
Board Interlock & 38,305,725 & 176,789,812 & 13,338,023 & 1,192,701,744 & 16,124,492 & 188,322,676 & 202\\
Scientific Collaboration & 30,861,046 & 189,969,041 & 20,067,279 & 637,689,557 & 28,203,800 & 284,255,735 &198\\
\bottomrule
\end{tabular}
\caption{The number of nodes $|N|$ and hyperedges $|E|$ in the higher-order network, followed by the number of links $|L|$, triangles $\tau$ and nodes in the largest connected component $\sigma$. Then follow the sum of the link weights $\Lambda$ in the projected network, and the number of different label values of the board interlock and of the scientific collaboration network.}
\label{tab:details}
\end{table}
}

\section{Results}
In this section, we show the results obtained by applying the methods presented in Section~\ref{sec:methods} on the two real-world network datasets introduced in Section~\ref{sec:data}.
In particular, in Section~\ref{subsec:overall_analysis} we present the insights obtained from the analysis of the considered measures of the overall network structure of both networks, whereas in Section~\ref{subsec:community_analysis}, we present the analysis on the contribution of hyperedges of different orders to the inter- and intra-label connections of the two considered networks.
\subsection{Contribution to overall network measures}
\label{subsec:overall_analysis}
The results we show in this section are related to how hyperedges of different orders contribute to the considered network measures that characterize the network structure at the micro, meso and macro level. 
\subsubsection{Link weights}
\label{subsubsec:weights_overall} 
\begin{figure}[ht]
    \centering
    \includegraphics[width=0.9\linewidth]{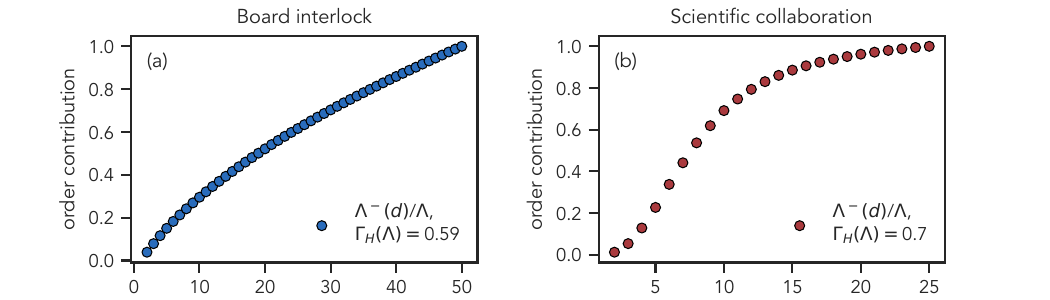}
    \caption{Order contribution $\Lambda^-(d)$ to \textbf{link weights}, normalized by the sum of the weights of links  $\Lambda$ in the projected network, as a function of the interaction order $d$ for board interlock (blue) and collaboration network (red).}
    \label{fig:order_relevance}
\end{figure}

We first show the contribution of hyperedges of different orders to the overall link weights in the two considered real-world networks.
The pairwise projected network is the most common representation in the literature to study the networks of board interlocks~\cite{heemskerk2013rise} and research collaborations~\cite{newman2001structure}. We quantify how directors with various number of appointments and scientific collaborations of varying size contribute to the sum of the link weights in this pairwise representation. 
In Figure~\ref{fig:order_relevance}, we can see that the order contribution $\Lambda (d)$ in the case of board interlock network increases much slower than in the scientific collaboration network, as the hyperedge order increases. Thus, in the board interlock network, the contribution to the link weights of the large order hyperlinks or "busy directors", i.e. directors appointed by a large set of different companies, is more evident than that of large research collaborations in the scientific collaboration network, as confirmed by the smaller order relevance measure for link weights obtained in the first network, when compared with the second.

\subsubsection{Triangles}
\label{subsubsec:triangles_overall} 

\begin{figure}[!b]
    \centering
    \includegraphics[width=0.85\linewidth]{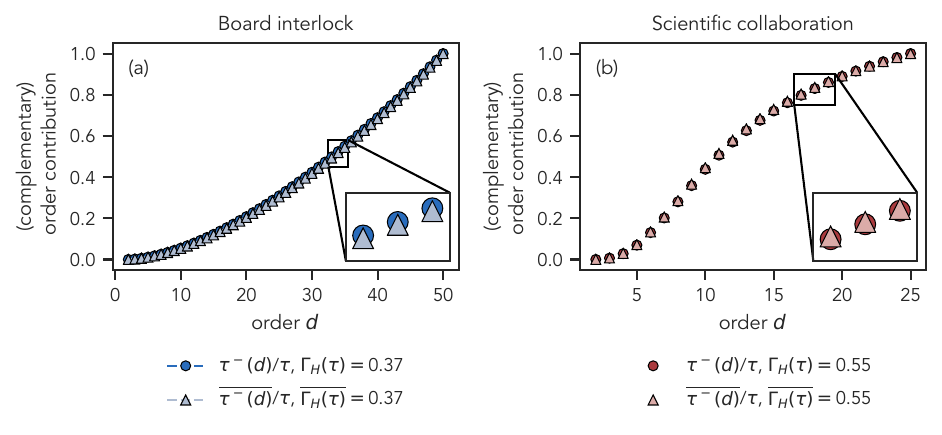}
    \vspace{-2em}
    \caption{Order contribution $\tau^-(d)$ (circles, darker color) to the \textbf{number of triangles} and its corresponding complementary $\overline{\tau^-(d)}$ (triangles, light color), normalized by the total number of triangles  $\tau$ in the projected network, as a function of the order $d$ for the board interlock (blue) and the collaboration network (red). }
    \label{fig:triangles}
\end{figure}

Moving beyond the local-scale perspective analyzed above, we then focus on a mesoscale network measure, the number of triangles. 
Previous work on scientific collaborations and board interlocks has shown how these networks usually have high values of the clustering coefficient.\cite{heemskerk2016corporate,newman2001structure} The interpretation of such results is however challenging, as with a pairwise representation, each collaboration or interlock involving more than $d>2$ nodes is decomposed into a clique, heavily increasing the number of triangles and thus potentially affecting clustering measures. We quantify such effects by computing the contribution of hyperlinks of different orders on the number of triangles, which, as mentioned in Section~\ref{subsubsec:triangles_def}, affects classic measures of global and local clustering. 
Figure~\ref{fig:triangles} shows the order contribution, its complementary and the corresponding order relevance for the board interlock and the collaboration networks. In general, we observe substantially smaller values of order relevance in both datasets: this suggests a more evident contribution of large-size hyperedges (busy directors and large scientific collaborations) to the number of triangles as compared to the link weights. Consistent with what was observed in Section~\ref{subsubsec:weights_overall}, the impact of busy directors is more evident in the board interlock network than that of large collaborations in the collaboration network, as indicated by the smaller value of order relevance $\Gamma_{H}(\tau)$.
From Figure~\ref{fig:triangles} we further observe that in both board interlock and collaboration networks $\Gamma_{H}(\tau) \approx \overline{\Gamma_{H}(\tau)}$, i.e., $\Delta_{H}(\tau) = 0$. 
This indicates that the contribution of each order is not influenced by which other order is also included, so it is neither synergistic nor redundant. This is likely due to the fact that the number of triangles is largely influenced by the projection of hyperedges with order $d\geq 3$ (closed triangles). As the number of triangles affects the two traditional network measures of clustering, these results suggest that caution should be taken in the interpretation of the large values of global/local clustering coefficients detected in board interlock and scientific collaborations, when these are represented as pairwise networks.

\label{subsubsec:connected_overall}
\begin{figure}[h]
    \centering
    \includegraphics[width=0.7\linewidth]{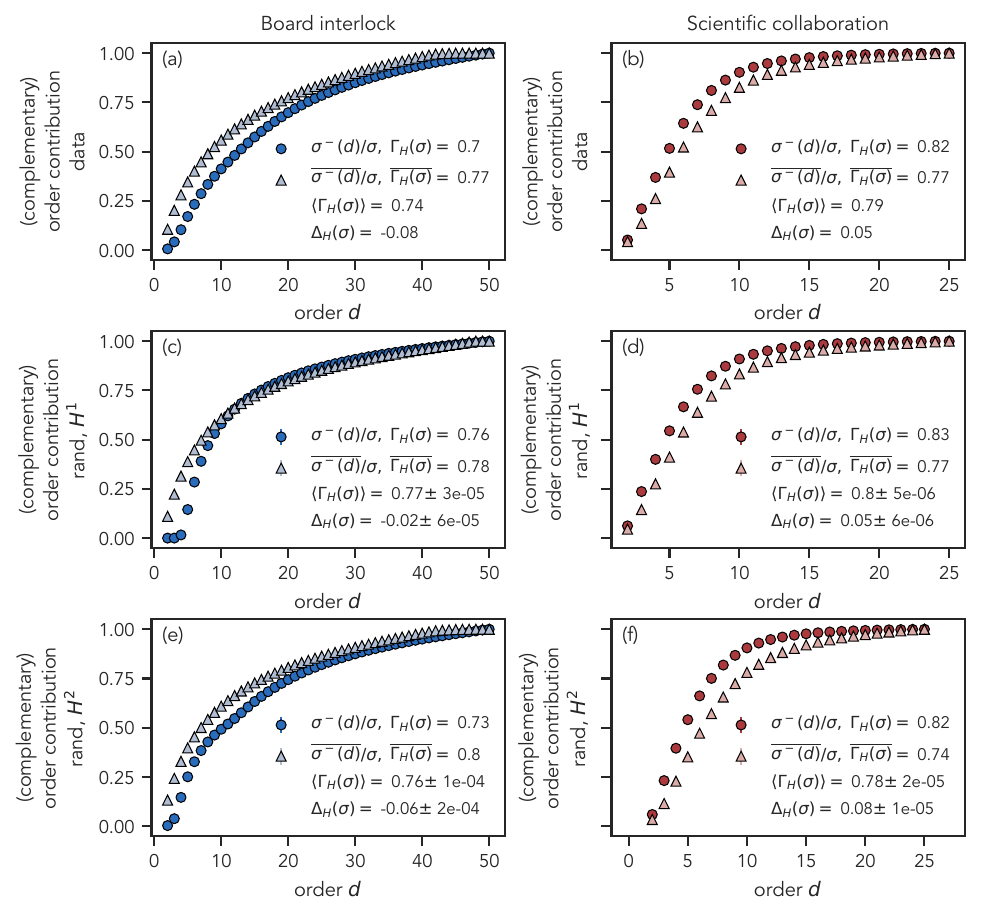}
    \caption{Order contribution $\sigma^-(d)$ (red) to the \textbf{largest connected component size} and its corresponding complementary contribution $\overline{\sigma^-(d)}$ (circle, light color), normalized by the largest connected component size $\sigma$ of the projected network, as a function of the order $d$ for the board interlock network (left, blue) and scientific collaboration network (right, red). The top row shows original network $H$ (a, b), the middle and bottom row are randomized networks: $H^1$ (c, d) and $H^2$ (e, f). 
    The randomized network $H^1$ preserves the number of hyperedges of each order attached to each node of the original network $H$, but not the group composition probability $P_{\ell}(d,k)$  of each label.
    Oppositely, the randomized network $H^2$ preserves the group composition probability $P_{\ell}(d,k)$ of each label of the original network $H$, but not the number of hyperedges of each order attached to each node. The values of the considered metrics obtained from the randomized null models $H^1$ and $H^2$ are averages over 10 independent realizations. The values of the average order relevance and order gap are reported together with their standard deviation. }
    \label{fig:connected_component_overall}
\end{figure}

\subsubsection{Connected Component}
After the local and mesoscale points of view presented above, here we quantify the contribution of hyperedges of different orders to 
the largest connected component.
The contribution of different orders in the context of board interlock network and scientific collaboration network can shed light on the specific roles of directors with different number of appointments or collaborations with different numbers of members in integrating the overall network. \cite{heemskerk2013rise}

Figures~\ref{fig:connected_component_overall}a and~\ref{fig:connected_component_overall}b show the order contribution to the largest component of the projected network $\sigma^-(d)$ and its complementary $\overline{\sigma^-(d)}$ .
By focusing only on the order contribution $\sigma^-(d)$ and the corresponding order relevance $\Gamma_{H}(\sigma)$, the contribution of busy directors to the size of the largest connected component seems more evident in the board interlock network than that of large scientific collaborations in the collaboration network, as shown by the smaller order relevance $\Gamma_{H}(\sigma)$ obtained for this network.
As discussed in Section~\ref{subsec:relevance}, $\sigma^-(d)$ and the corresponding order relevance $\Gamma_{H}(\sigma)$ are defined by progressively including hyperedges from the smallest to the largest order. However, the order contribution (and its corresponding order relevance can be influenced by the way in which hyperedges of different orders are progressively included, i.e., from smallest to largest order or vice-versa.
The values of the complementary order relevance $\overline{\Gamma_{H}(\sigma)}$ show that the contribution of large orders to the largest connected component seems similar in the two datasets.
Overall, we thus observe a smaller average order relevance in the board interlock network as compared to the scientific collaboration network. This means that busy directors contribute more to integrating the board interlock network than large collaborations to the scientific collaboration network. These results are consistent with what we observed in Sections~\ref{subsubsec:weights_overall} and~\ref{subsubsec:triangles_overall}.

Moreover, the order relevance gap $\Delta_{H}(\sigma)$ is smaller than 0 in the board interlock, and larger than 0 in scientific collaboration network.
This suggests a substantial difference in the structural organization of the network: in  the board interlock network, the contribution of directors with many (few) appointments is larger if directors with few (many) appointments are also included, suggesting a synergistic contribution of different orders.
Thus, in this network, the integration of companies into the largest connected component is the result of a combined contribution of the directors connecting large and those connecting a small number of companies. 
If either the directors connecting many companies or those connecting few of them were absent, the contribution of the remaining directors would be reduced.
This is different from the scientific collaboration network, where the contribution of high-order hyperedges is reduced if lower-order interactions are also included, suggesting a redundant contribution of hyperedges of different orders. This means that, in the scientific collaboration network, part of the nodes integrated in the largest connected components by large collaborations could be integrated just by the presence of small collaborations, if large collaborations were absent. The same would also apply to the contribution of large collaborations if small ones were absent.
Next, we investigate how two key network properties influence the value of the average order relevance, the order relevance gap and the corresponding synergistic and redundant contribution of hyperedges with different orders. 
This is done by comparing the values obtained from real data with the randomized reference models introduced in Section~\ref{subsec:random_ref}. 

The first considered network property is the number of hyperedges at each order attached to each node. In the board interlocks, these randomizations preserves the number of appointments of each director sitting in the board of a given company/node.
Note that the results obtained from randomizations shown in the remainder of this paper are averages over 10 independent realizations, reported together with the corresponding standard deviations. 
Regarding board interlock network (see Figure~\ref{fig:connected_component_overall}c), the value of the average order relevance is slightly higher, while the positive value of $\Delta_{H}(\sigma)$ is reduced by the randomization $H^1$, which preserve the number of hyperedges of each size attached to each node (as discussed in Section~\ref{subsec:random_ref}). 
Such differences suggest that this network property alone reduces the contribution of busy directors and cannot reproduce the synergistic contribution of directors with different numbers of appointments observed in the order contribution to the largest connected component observed in the real board interlock network.
However, in the considered networks, each node is assigned a label, and nodes with the same or different labels could be connected more likely by specific orders. Such differences are captured by the group composition probability and the group balance of each label, as discussed in Section~\ref{subsec:label_group_def}. As in both considered networks,  labels correspond to geographical locations, we refer to the label composition of each hyperlink, i.e., the number of nodes of each label connected by each hyperlink, as its international composition.
Similarly to the case of $H^1$, the average order relevance of the randomized network $H^2$ is slightly increased.
Differently from randomization $H^1$, however, the value of the order gap is only slightly reduced by randomization $H^2$ as shown in Figure~\ref{fig:connected_component_overall}e).
These two observations suggest that preserving the international composition of hyperedges, i.e., how companies of the same or different countries are interlocked by directors with varying number of appointments, can alone reproduce a similar value of synergy, but still reduce the contribution of busy directors (large hyperlinks). 
On the other hand, for the collaboration network, we obtain substantially different results. Both randomizations $H^1$ (Figure~\ref{fig:connected_component_overall}d) and $H^2$ (Figure~\ref{fig:connected_component_overall}f) almost preserve the average order relevance observed in Figure~\ref{fig:connected_component_overall}b. Thus, the two properties preserved in the corresponding randomized null models $H^1$ (number of hyperedges of each order attached to each node) and $H^2$ (label composition of each hyperedge) can reproduce the contribution of large collaborations to the largest connected component. Moreover, randomization $H^1$ (Figure~\ref{fig:connected_component_overall}d) slightly increases the negative value of $\Delta_{H}(\sigma)$ observed in Figure~\ref{fig:connected_component_overall}b: by preserving only the number of collaborations/hyperlinks of each size that a node/researcher was involved in, we can already preserve the redundancy observed in the real network.
The value of order gap for randomization $H^2$ (Figure~\ref{fig:connected_component_overall}f) is instead larger than that observed in the real network $H$ (Figure~\ref{fig:connected_component_overall}b). In this case, preserving the international compositions of each research collaboration, without the number of collaborations of each size that the node/researcher was involved in corresponds to a much higher redundancy than what is observed in the real network $H$.
In the scientific collaboration network, only the sequence of collaborations of each size each researcher was involved in can reproduce similar values of redundancy observed in the real network (Figure~\ref{fig:connected_component_overall}b), while preserving the international composition of each research collaboration alone results in more redundancy. This means that the redundancy observed in the collaboration networks seems to be due to individual behavior of researchers to engage in collaborations of specific sizes rather than to their tendency to collaborate with collaborators based in specific countries.

\subsection{Contribution to intra-/inter-label ties}
\label{subsec:community_analysis}
In Section~\ref{subsec:overall_analysis} we have shown how hyperedges of different orders contribute to the considered properties of the overall network. Moreover, in Section~\ref{subsubsec:connected_overall}, we observe that the synergistic contribution of orders observed in the board interlock network appears to be partially explained by the label composition of hyperedges, characterized by the group composition probability of the label $\ell$, i.e., $P_{\ell}(d,k)$.
In this section, we thus investigate the patterns of connections among nodes with same/different labels of the 50 labels with the largest number of intra- and inter- label pair connections, i.e., $\Lambda_{\ell,intra}$ and $\Lambda_{\ell,inter}$\footnote{We excluded labels with an unbalanced number of (inter-)intra-label connections by ranking each label by the smallest of the values between $\Lambda_{\ell,intra}$ and $\Lambda_{\ell,inter}$, and then taking the top 50 labels.}. We apply the methods presented in Section~\ref{subsec: Community}, to study how hyperedges (directors/scientific collaborations) of different sizes connect nodes (directors/researchers) with the same or different labels (countries) in pairs (Section~\ref{subsec:inter_intra_pair}) or groups (Section~\ref{subsec:inter_intra_group}), and how nodes with the same labels are integrated in the largest connected component of the two networks (Section~\ref{subsec:inter_intra_component}). 

\subsubsection{Intra- and inter- label pairwise ties}
\begin{figure}[h]
    \centering
    \includegraphics[width=0.8\linewidth]{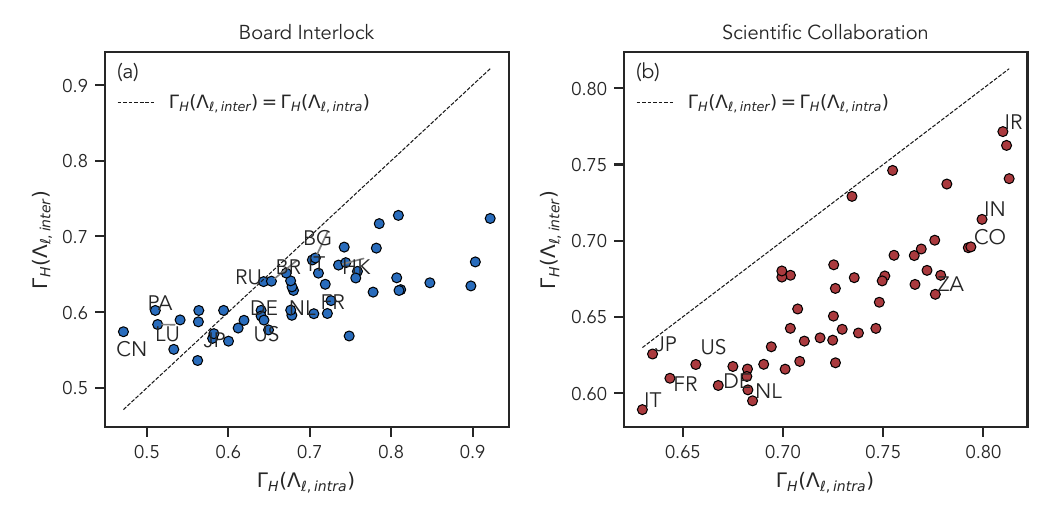}
    \caption{Order relevance of \textbf{the sum of the national link weigths} ($\Lambda^{-}_{\ell,intra}(d)$, horizontal axis) versus the order relevance of \textbf{the sum of the international link weights} ($\Lambda^{-}_{\ell,inter}(d)$, vertical axis) for each considered country (denoted using the ISO2 country code) in the board interlock (a) and scientific collaboration (b) network.}
    \label{fig:enter-label}
\end{figure}

We first investigate how the hyperedges of different sizes contribute to the strength of the intra-label and inter-label pairwise connections.
We compute the order relevance of the sum of weights for intra-label links and inter-label links, that is, $\Lambda^{-}_{\ell,intra}(d)$ and $\Lambda^{-}_{\ell,inter}(d)$. 
As discussed in Section~\ref{sec:data}, in both board interlock and scientific collaboration data, labels correspond to countries; we will refer to intra-label links as "national" ones, while inter-label links as "international" ones.
In Figure~\ref{fig:enter-label}, we observe that, in both datasets,
the order relevance of the sum of the weights of the national links is greater than the corresponding sum of the weights of the international ones in the vast majority of countries, suggesting an overall larger contribution of hyperedges with larger size to the weights of international links rather than to national ones. This holds for all countries considered in the collaboration network, while in the board interlock network, we also observe a small set of countries, such as China or Panama, where $\Lambda^{-}_{\ell,intra}(d) < \Lambda^{-}_{\ell,inter}(d)$: in these countries, large orders contribute more to national connections than to the international ones.  

\label{subsec:inter_intra_pair}
\subsubsection{Intra- vs inter-label group ties}

Directors or scientific collaborations connect sets of companies or researchers that can be larger than pairs. In this section, we investigate how hyperedges in these networks connect companies/researchers based in different countries, by computing for each country the group composition probability $P_{\ell}(d,k)$ and the group balance $\Phi_{\ell}$ introduced in Section~\ref{subsec:inter_intra_group}. Detailed comparisons with the two randomized null models are provided in the Supplementary Material. 

In Figure~\ref{fig:board_overlap_label}a we show the value of the group balance $\Phi_{\ell}$ for each considered country in the board interlock network.
Four different observations can be obtained from this figure.
First, we see small values of group balance in BRIC countries (Brazil, Russia, India, and China) and Japan. This indicates that, in these cases, directors tend to connect companies based in the country with large groups of other companies based in the same country.
Moreover, the highest values of group balance are reached by U.S., Hong Kong and Luxemburg: in these cases, directors tend to connect few companies based in the considered country with many other companies based in other countries. Similar differences in the values of the label group balance are also reflected in the different group composition distributions $P_{\ell}(d,k)$ and the corresponding conditional average $\langle k_{\ell} \rangle_d$ of China (low group balance, Figure~\ref{fig:board_overlap_label}b) and Luxemburg (high group balance, Figure~\ref{fig:board_overlap_label}c). Finally, the tendencies of directors to connect companies based mainly in the same or different countries observed in the two discussed groups of countries are rather consistent across the overall number of director appointments (hyperedge order) $d$.

\begin{figure}[!b]
    \centering
    \includegraphics[width=0.7\linewidth]{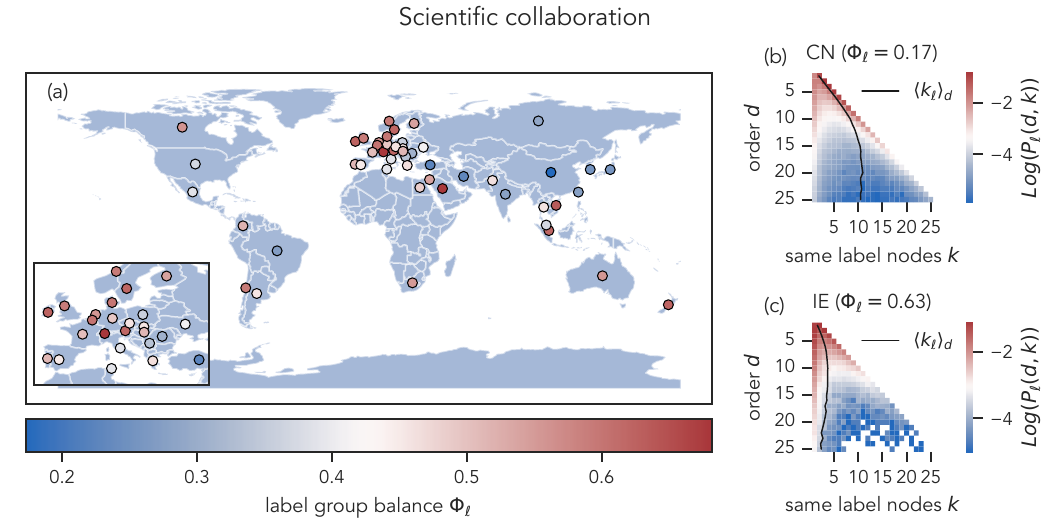}
    \caption{\textbf{Group balance} $\Phi_{\ell}$ of each country in the \textbf{scientific collaboration network} (a), the country's group composition distribution $P_{\ell}(d,k)$ together with the average number of nodes $\langle k_{\ell} \rangle_d$  with same label connected by a hyperedge of order $d$ for China (b) and Ireland (c).}
    \label{fig:collab_overlap_label}
\end{figure}

We obtain slightly different results for the collaboration network.
While BRIC countries also tend to have a low group balance, we generally observe a lower diversity of group balance values across country/labels, as shown by the larger minimum value and the smaller maximum value of group balance in Figure~\ref{fig:collab_overlap_label}a. By looking at the group composition probability and the corresponding conditional average conditional $\langle k_{\ell} \rangle_d$ of countries with low (e.g., China, Figure~\ref{fig:collab_overlap_label}b) and high (e.g., Ireland, Figure~\ref{fig:collab_overlap_label}c) group balance, we further observe that, in general, large collaborations usually involve a smaller ratio of national collaborators than those involving many members.
Despite the limitations discussed in section~\ref{subsec:label_group_def}, the label group balance seems to capture the national differences of the group composition probability observed in both considered networks relatively well. 

Apart from noteworthy countries discussed above, we also performed a more systematic comparison of country-specific results for group balance, and known World Bank indicators of both the country's wealth and its involvement in international trade. In both networks, we observe that the countries with a high group balance are in general those with a higher Gross Domestic Product (GDP) per capita (board interlock: \textit{r}=0.45, \textit{p}=0.001, scientific collaboration: \textit{r}=0.55, \textit{p}<0.001). In the board interlock network, the countries with a high group balance are also those with a higher inclination to international trade (\textit{r}=0.48, \textit{p}<0.001). The Supplementary Material presents more detailed results of this analysis.


\label{subsec:inter_intra_group}
\begin{figure}[!b]
    \centering
    \includegraphics[width=0.7\linewidth]{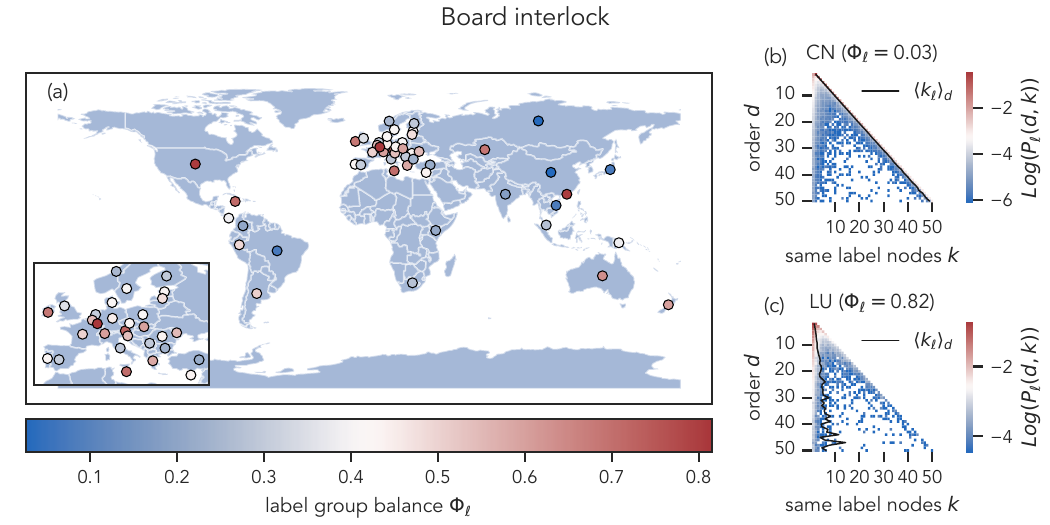}
    \caption{\textbf{Group balance} $\Phi_{\ell}$ of each country in the \textbf{board interlock network} (a), the country's group composition distribution $P_{\ell}(d,k)$ together with the average number of nodes $\langle k_{\ell} \rangle_d$ with same label connected by a hyperedge of order $d$ for China (b) and Luxembourg (c).}
    \label{fig:board_overlap_label}
\end{figure}

\subsubsection{Node labels in the largest connected component}
\label{subsec:inter_intra_component}
\begin{figure}[!b]
    \centering
    \includegraphics[width=0.82\linewidth]{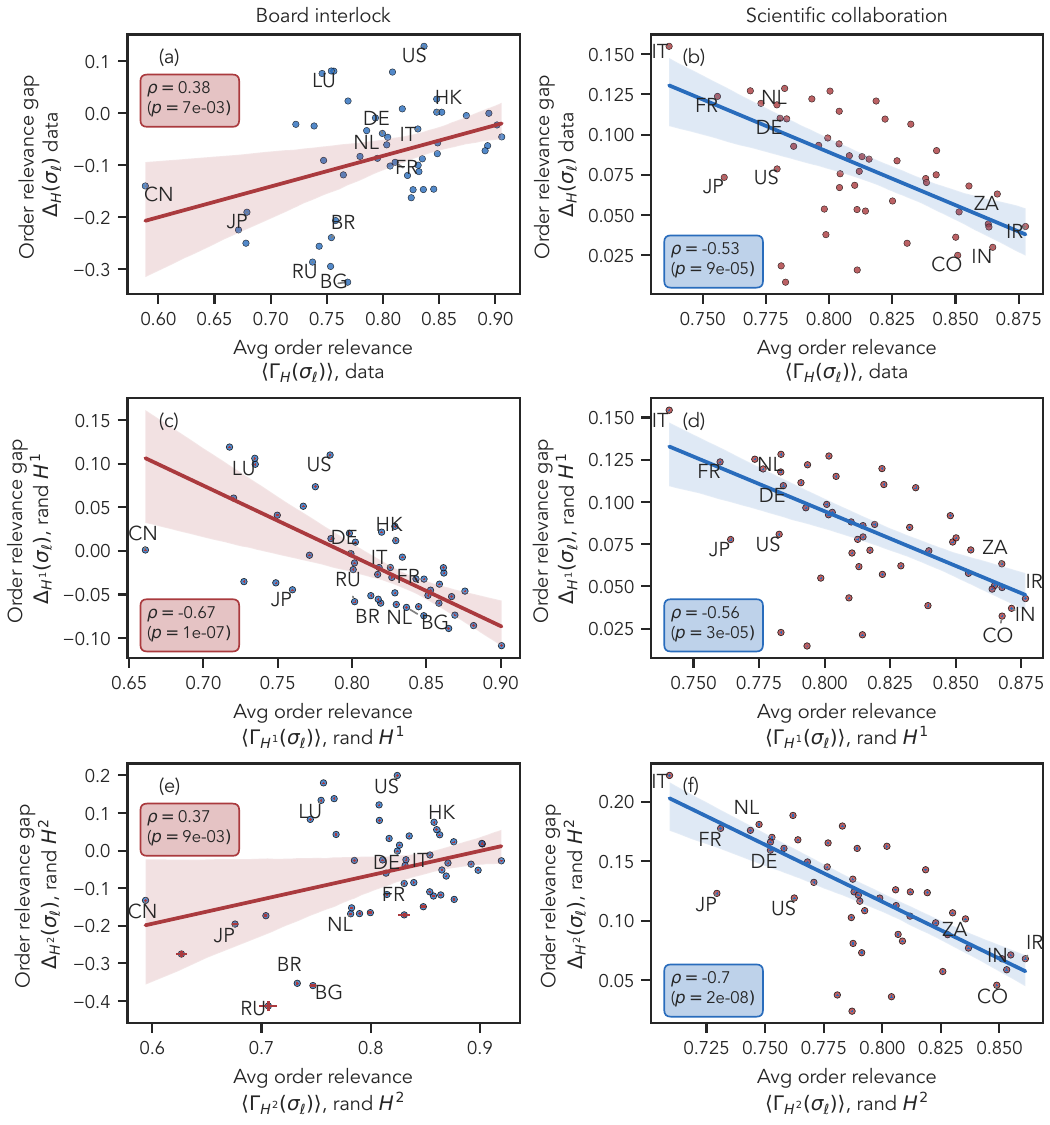}
    \caption{\textbf{Scatter plot of the average order relevance $\langle\Gamma_{H}(\sigma_{\ell}\rangle$ vs the order relevance gap $\Delta_{H}(\sigma_{\ell})$} of the number of nodes of each country in the largest connected component for the \textbf{network $H$} (c), and its \textbf{two randomised networks: $H^1$} (d) and \textbf{$H^2$} (e) for board interlock (left) and scientific collaboration (right). The randomized network $H^1$ preserves the number of hyperedges of each order attached to each node of the original network $H$, but not the group composition probability $P_{\ell}(d,k)$  of each label. 
    Oppositely, the randomized network $H^2$ preserves the group composition probability $P_{\ell}(d,k)$ of each label of the original network $H$, but not the number of hyperedges of each order attached to each node. The order gap and average order relevance of the two randomized null models $H^1$ and $H^2$ are the average, together with the relative standard deviations obtained from 10 independent realizations. We further show the linear fit and its corresponding 95\% confidence intervals, together with the value of the linear correlation coefficient $\rho$ and corresponding $p$-values.}
    \label{fig:local_component_scatter}
\end{figure}

After investigating the contribution of directors and collaborations of varying size to the national and international pairwise connections and the international composition of group ties, here we evaluate how collaborations and directors integrate nodes from different countries in the largest connected components of the corresponding projected networks. As discussed in Section~\ref{subsec:label_component_def}, we quantified this using $\sigma_{\ell}$.
Figure~~\ref{fig:local_component_scatter}a shows the average order relevance $\langle \Gamma_{H}(\sigma_{\ell}) \rangle$ and the order gap  $\Delta_{H}(\sigma_{\ell})$ for each considered country in the board interlock network.
Overall, we observe general high values of the average order relevance in the US and in the majority of Western European countries, while we see relatively lower values in Bulgaria, Brazil, Russia and China. Moreover, a variety of different order gap values across the considered countries is observed, ranging from more synergistic contribution of orders, i.e., $\Delta_{H}(\sigma_{\ell}) < 0$, in countries such as Bulgaria, Russia, Brazil and China, to those with a redundant contribution, such as United States.
Notably, countries with relatively lower average order relevance tend to show a more negative order gap, while those with a lower average order relevance are more likely to have a higher order gap value, as shown by the moderate positive correlation between the two measures (see Figure~\ref{fig:local_component_scatter}a). In the board interlock network, thus, the countries integrated in the global network by a stronger contribution from directors with many appointments are also those where this contribution is synergistic with that of directors with a smaller number of appointments. 
Interestingly, a comparison with World Bank indicators shows that countries with a higher order relevance gap seem to be richer and more open to international trade than those with a lower order relevance gap value. This suggests that companies with higher values are more likely to be integrated in the overall network by a more redundant contribution of the interlocking directors (see Supplementary Materials for details).

In Section~\ref{subsubsec:connected_overall}, we have shown that in the board interlock network, both randomizations  $H^1$ and $H^2$ slightly reduce the average order relevance of the overall largest connected component, but the overall synergistic contribution of different orders to this network measure is reduced only when randomization $H^1$ is applied. This randomization preserves the number of hyperedges of different orders attached to each node, but not the label group composition of the hyperedges.
As shown in Figure~\ref{fig:local_component_scatter}c, by applying randomization $H^1$, the diversity of the values of the average order relevance and the order gap observed are reduced in $H^1$ (see detailed results in Supplementary Material). Moreover, a negative correlation between the average order contribution and the order gap is observed, different from the positive one observed in $H$ (Figure~\ref{fig:local_component_scatter}a). Randomization $H^1$ cannot reproduce the diversity in local synergistic and redundant patterns observed in $H$, nor the positive correlation between the values of these two measures observed in the board interlock network. 
These two properties are instead observed in randomization $H^2$ (Figure~\ref{fig:local_component_scatter}e), which preserves the group composition probability of each country. These properties seem thus related to the national differences with which companies based in a country interlock with companies in the same or different country by directors.

Differently, in the scientific collaboration network, higher values of average order relevance (Figure~\ref{fig:local_component_scatter}b) and no synergistic contribution (Figure~\ref{fig:local_component_scatter}b) are observed also at the country level. Instead, for any country, we observe a redundant contribution of orders ($\Delta_{H}(\sigma_{\ell})>0$). We further observe relatively smaller values of average order relevance in Western Europe, US and Japan, while large values are instead observed in Colombia, Iran, South Africa and India. Regarding the order gap, large redundancy values (high order gap) are especially observed in Europe, while the lowest value of the order relevance gap is observed for China. We further observe a negative correlation between the average order contribution and the order gap, as shown in Figure~\ref{fig:local_component_scatter}b.
Differently from the case of the board interlock network, in scientific collaborations, countries integrated in the overall network by a larger contribution of large collaborations contribute the most to integrating the local national community to the global scientific network are more likely those in which this contribution is more redundant with that of collaboration of smaller size.

Moreover, in Figure~\ref{fig:local_component_scatter}d we observe that randomization $H^1$ does not significantly change the results discussed for the real network, while randomization $H^2$ increases redundancy at the local level (see Figure~\ref{fig:local_component_scatter}f).
In the scientific collaboration network, redundancy thus seems mainly the result of the individual behavior of the researchers to engage in collaborations of different sizes, rather than their preference in collaborating with researchers based in specific countries.
This is consistent with the conclusions obtained by comparing the values of the order relevance gap of the largest connected component $\sigma$ in the real network $H$ and in the randomized models $H^1$ and $H^2$ in the overall analysis of Section~\ref{subsubsec:connected_overall}.

\section{Discussion and Conclusion}
In this paper, we presented a methodology to precisely quantify the contribution of hyperedges of different orders to local, mesoscale and global properties of the overall network, and to the group and pairwise interactions among nodes belonging to the same or different partition based on a meaningful node label. 
We first proposed the \textit{order contribution} to quantify the contribution of hyperedges of different size to a given network measure, and then the \textit{order relevance} to assess the relative contribution of large and small order hyperedges to measures at the local, meso and global scale. 
Moreover, we proposed a measure to assess the tendency of hyperedges of different orders to connect nodes with either the same or different labels via the label \textit{group balance}. 
Finally, two randomized null models are used, taking into account the number of hyperedges of different orders attached to each node and the labels of nodes connected by each hyperedge.

We applied our methods to two different networks, namely a board interlock and a scientific collaboration network, where node labels indicate the geographic location associated with the node.
Our analysis shows that, overall, the contribution of large orders at the local, mesoscale, and global level is more evident in the board interlock than in the collaboration network.
Moreover, in the board interlock network, the contribution to the largest connected component of large orders is larger when also small orders are included, i.e., there is "synergistic" contribution of orders to the emergence of the largest connected component. 
Differently, in the scientific collaboration network, the contribution to the largest connected component of large orders is smaller when also small orders are included, suggesting a "redundant" contribution of different orders.
The synergistic contribution observed in the board interlock network is much less evident in the randomized null model that alters the tendency of companies at a country to be interlocked with companies based on the same or different countries. Differently, the magnitude of redundant contribution in the scientific collaboration network is not reproduced by the randomized null models that cannot preserve the individual tendency of researchers to participate in collaborations of different sizes.
Taking a global perspective and investigating the difference between national and international connections, we show that, generally, larger hyperedges contribute more to the international pairwise connections than to national ones in both network datasets. 
Substantial differences across labels/countries are however observed in terms of how hyperedges of different sizes connect nodes with the same label, reflecting differences in wealth and inclinations to international trade of different countries. For example, in BRIC countries, we observe a tendency for directors to link companies with the same country. Oppositely, in the US, Luxembourg, and Hong Kong, they tend to connect a relatively small number of companies in that country with companies based in other countries, creating transnational ties.
Finally, in the board interlock network, we observe that countries with higher synergy are usually those where the contribution of larger orders is more evident. Conversely, countries where the contribution of large orders is less evident tend also to have lower synergy, or even a redundant contribution. 
In the scientific collaboration network, the overall contribution of large orders in integrating each country in the global network is much lower, and only redundant contributions were detected, with higher values of redundancy associated with countries with a more evident contribution from large collaborations.
Such differences seem to reflect the differences in national wealth (both networks), and academic freedom among the considered countries.

Our proposed methods allow practitioners from different fields (e.g. social sciences, biology, medicine, or ecology) to obtain meaningful insights from the higher-order analysis of real world network data.
A promising direction for future research is to generalize these methods to temporal or directed higher-order networks and to enlarge the set of considered topological network metrics to obtain further insights into the impact of higher-order connections on network topology. 

\section*{Declarations}

\subsection*{Ethics approval and consent to participate}
Not applicable.

\subsection*{Consent for publication}
I confirm that I understand EPJ Data Science is an open access journal that levies an article processing charge per articles accepted for publication. By submitting my article, I agree to pay this charge in full if my article is accepted for publication.

\subsection*{Availability of data and materials}
The data that support the findings of this study are available from the CORPNET project and CWTS, but restrictions apply to the availability of these data, which were used under license for the current study, and so are not publicly available. Data are however available from the authors upon reasonable request and with permission of CORPNET consortium and CWTS.
The code is publicly available at  \url{https://github.com/aceria/higher_order_relevance}. 

\subsection*{Competing interests}
The authors declare that they have no competing interests.

\subsection*{Funding}
Not applicable.

\subsection*{Authors' contributions}
AC and FT conceived and designed the study, read, improved and approved the final manuscript.
AC performed the analysis and wrote the manuscript.

\subsection*{Acknowledgments}

Authors are grateful to Eelke Heemskerk and by extension the CORPNET project (\url{https://corpnet.uva.nl}) for providing access to the corporate board interlock data. 
Authors are also thankful to Hanjo Boekhout and the Centre for Science and Technology Studies (\url{https://www.cwts.nl}) for providing access to the preprocessed anonymized version of the scientific collaboration data.
Authors further aknowledge Leonardo Di Gaetano for useful comments and Dyliara Valeeva for providing information about the board interlock network. Finally, authors thank the anonymous reviewers for their helpful suggestions. 

\bibliography{sample}

\include{supplementary_material}

\end{document}

%% file: supplementary_material.tex
\appendix
\setcounter{figure}{0}
\setcounter{table}{0}
\renewcommand{\figurename}{}
\renewcommand{\thefigure}{SM\arabic{figure}}
\renewcommand{\tablename}{}
\renewcommand{\thetable}{SM\arabic{table}}







\section*{Supplementary Material}

\begin{table}[h!]
    \centering
    \resizebox{0.95\textwidth}{!}{%
    \begin{tabular}{|c|c|c|}
        \hline
        \textbf{Symbol} & \textbf{Name} & \textbf{Definition}\\
        \hline
        \multicolumn{3}{|c|}{\textbf{General definitions of higher-order networks}}\\
        \hline
        \(H = (N,E)\) & Hypergraph & Hypergraph with node set \(N\) and hyperedge set \(E\)\\
        \hline
        \(e =\{v_1,\dots,v_d\} \) & Hyperedge & Connection among nodes \(\{v_1,\dots,v_d\}\) \\
        \hline
        \(d \) & Order & Number of nodes connected by a hyperedge\\
        \hline
        \( G = (N,L) \) & Pairwise (or projected) network & \parbox{8cm}{Network with node set \(N\) and link set \(L\). A pair of nodes is connected by a link in \(G\) if the nodes are connected by a hyperedge of any order in \(H\)}\\
        \hline
        \( w(v_i,v_j) \) & Weight of a link in the pairwise network & Number of different hyperedges connecting \(v_i\) and \(v_j\)\\
        \hline
        \multicolumn{3}{|c|}{\textbf{Definitions of order contribution hypergraphs}}\\
        \hline
        \( H^{-}(d) \) & \(d\)-order contribution hypergraph & \parbox{8cm}{Sub-hypergraph of \(H\), with same node set \(N\) and hyperlinks with order \(d' \leq d\)}\\
        \hline
        \( H^{+}(d) \) & Inverse \(d\)-order contribution hypergraph & \parbox{8cm}{Sub-hypergraph of \(H\), with same node set \(N\) and hyperlinks with order \(d' > d\)}\\
        \hline
        \( M^-(d) \) & Order contribution & \parbox{8cm}{Value of the network measure \(M\) computed on the \(d\)-order contribution hypergraph \(H^{-}(d)\)}\\
        \hline
        \( M^+(d) \) & Inverse order contribution & \parbox{8cm}{Value of the network measure \(M\) computed on the inverse \(d\)-order hypergraph \(H^{+}(d)\)}\\
        \hline
        \( \overline{M^-(d)} \) & Complementary order contribution & \( \overline{M^-(d)}= M_{max} -  M^+(d) \)\\
        \hline
        \multicolumn{3}{|c|}{\textbf{Definitions of order relevance measures}}\\
        \hline
        \( \Gamma_{H}(M) \) & Order relevance & \( \Gamma_{H}(M) = \frac{\sum_{d = d_{min}} ^ {d_{max}} (M^{-}(d)/M_{max})  - 1}{d_{max} - d_{min}} \)\\
        \hline
        \( \overline{\Gamma_{H}(M)} \) & Complementary order relevance & \( \overline{\Gamma_{H}(M)} = \frac{\sum_{d = d_{min}} ^ {d_{max}} (\overline{M^-(d)}/M_{max})  - 1}{d_{max} - d_{min}} \)\\
        \hline
        \( \Delta_{H}(M) \) & Order relevance gap & \parbox{8cm}{\( \Delta_{H}(M) = \Gamma_{H}(M) - \overline{\Gamma_{H}(M)} \)
        \[
        \begin{cases}
        >0 & \text{redundancy} \\
        <0 & \text{synergy}
        \end{cases}
        \]}\\
        \hline
        \( \langle \Gamma_{H}(M) \rangle \) & Average order relevance & \( \langle \Gamma_{H}(M) \rangle = \frac{\Gamma_{H}(M) + \overline{\Gamma_{H}(M)}}{2} \)\\
        \hline
        \multicolumn{3}{|c|}{\textbf{Definitions of topological network measures}}\\
        \hline
        \( \Lambda \) & Overall sum of link weights & Sum of link weights in the pairwise projected network \(G\)\\
        \hline
        \( \tau \)  & Number of triangles & Total number of triangles in the pairwise projected network \(G\)\\
        \hline
        \( \sigma \) & Size of the largest connected component & \parbox{8cm}{Number of nodes in the largest connected component of the pairwise projected network \(G\)}\\
        \hline
        \multicolumn{3}{|c|}{\textbf{Definitions of intra-/inter-label measures}}\\
        \hline
        \( \Lambda_{\ell,intra} \) & Sum of the intra-label link weights of label \(\ell\) & \parbox{8cm}{Sum of the weights of links connecting pairs of nodes with the same label \(\ell\) in the pairwise projected network \(G\)}\\
        \hline
        \( \Lambda_{\ell,inter} \) & Sum of the inter-label link weights of label \(\ell\) & \parbox{8cm}{Sum of the weights of links connecting a node with label \(\ell\) to a node with label \(\ell'\neq\ell\) in the pairwise projected network \(G\)}\\
        \hline
        \( P_{\ell}(d,k) \) & Group composition probability of label \(\ell\) & \parbox{8cm}{Probability that a random hyperlink connects \(d\) nodes, \(k\) of which have label \(\ell\)}\\
        \hline
        \( \Phi_{\ell} \) & Group balance of label \(\ell\) & \( \Phi_{\ell} = \frac{\sum_{d}\sum_kP_{\ell}(d,k)\  (d-k)}{\sum_d\sum_kP_{\ell}(d,k)\ (d-1)} \)\\
        \hline
        \( \sigma_{\ell} \) & Nodes with label \(\ell\) in the largest connected component & \parbox{8cm}{Number of nodes with label \(\ell\) in the largest connected component of the pairwise projected network \(G\)}\\
        \hline
    \end{tabular}%
    }
    \caption{General definitions, order contribution hypergraphs, order relevance measures, and topological network measures.}
    \label{tab:merged_metrics}
\end{table}

\subsection*{Derivation of the bounds of the order relevance}
In this section, we will show that the order relevance is bounded between 0 and 1 for monotonically increasing functions, and we show a slightly modified formulation to study monotonically decreasing functions.
\subsubsection*{Monotonically increasing order contribution}
As defined in Section 2.2 of the main manuscript, the order relevance of the monotonically increasing network measure $M$ of a higher-order network $H$ is 
\begin{equation}
    \Gamma_{H}(M) = \frac{(\sum_{d=d_{min}}^{d_{max}}M^-(d)/M_{max})-1}{d_{max} - d_{min}} 
\end{equation}

To derive this expression, we start by introducing the sum over all orders of the order contribution $M^-(d)$, i.e. $\sum_{d=d_{min}}^{d_{max}}M^-(d)$. Two extreme cases bound this quantity, the one in which only the smallest order contributes to $M$, and the one in which only the largest order $d_{max}$ contributes. The order contribution $M^-(d)$ is a monotonically increasing function of the order $d$ or 
a constant. The order contribution is obtained by measuring the value of the network measure $M$ while progressively including larger order hyperlinks. Thus the case in which only the smallest order $d_{min}$ contributes is the constant case $M^-(d) = M_{max}$ for each order $d$, i.e., the maximum value $M_{max}$ is already obtained by including only the smallest order $d_min$. The opposite case is instead the one in which only the largest order $d_{max}$ contributes to $M$, meaning that the order contribution is null for any order $d'\neq d$ and is equal to $M_{max}$ when $d = d_{max}$, that is, $M^-(d)= M_{max}\ \delta_{d,d_{max}}$, where $\delta_{i,j}$ is the Kroenecker delta.
By substituting the values of the two cases in the sum over all orders of the order contribution we get
\begin{equation}
    \sum_{d=d_{min}}^{d_{max}}M_{max}\ \delta_{d,d_{max}} \leq \sum_{d=d_{min}}^{d_{max}}M^-(d) \leq  \sum_{d=d_{min}}^{d_{max}}M_{max}
\end{equation}
 After explicitly computing the two bounds, we obtain the following:

\begin{equation}
    M_{max} \leq \sum_{d=d_{min}}^{d_{max}}M^-(d) \leq M_{max} (d_{max} - d_{min}+1)
\end{equation}
From this expression, we normalize each of its parts by $M_{max}$, then we subtract to each part 1, obtaining
\begin{equation}
    0 \leq \frac{(\sum_{d=d_{min}}^{d_{max}}M^-(d)/M_{max})-1}{d_{max} - d_{min}} \leq  1
\end{equation}

\subsubsection*{Monotonically decreasing order contribution}
In this second subsection, we propose a modified definition of the order relevance to study the contribution of small and large orders to a monotonically decreasing order contribution. 
The first step to compute the order relevance of a monotonically decreasing order contribution $M_{H}^-(d)$ is subtracting its minimum value $M_{min}$, obtaining the centered order contribution $\tilde{M}^-(d)$ = $M^-(d)-M_{min}$ and its corresponding maximum value $\tilde{M}_{max} = M_{max} - M_{min}$.
Similarly to what was discussed in the previous section, we introduce the sum over all the orders of the centered order contribution $\tilde{M}^-(d)$, i.e. $\sum_{d=d_{min}}^{d_{max}}\tilde{M}^-(d)$. 
In the case of constant order contribution, $\tilde{M}^-(d) = 0$ for any $d$, so the value of the sum of the centered order contribution is $\sum_{d=d_{min}}^{d_{max}}\tilde{M}^-(d) = 0$. 
Then, the smallest possible value of the sum over all the orders of the order contribution $\sum_{d=d_{min}}^{d_{max}}\tilde{M}^-(d)$ is 0. In contrast, the highest value is obtained when the order contribution is $\tilde{M}^-(d) = \tilde{M}_{max}$ for any $d<d_{max}$ and $\tilde{M}^-(d) = 0$ for $d=d_{max}$, that is, $\tilde{M}^-(d) = \tilde{M}_{max}(1-\delta_{d,d_{max}})$, with $M_{max}\neq M_{min}$.

As we already showed that in the case of constant order contribution, the sum is equal to 0, we can compute the sum over all the orders of the order contribution in the case that the order contribution is a monotonic, non-constant decreasing function, i.e. $\tilde{M}_{max}\neq0$, we obtain

\begin{equation}
    0 < \sum_{d=d_{min}}^{d_{max}}\tilde{M}^-(d) \leq  \sum_{d=d_{min}}^{d_{max}}\tilde{M}_{max}(1-\delta_{d,d_{max}})
\end{equation}
Note that, in case of monotonically decreasing order contribution, we obtain the smallest value of the sum $\sum_{d=d_{min}}^{d_{max}}\tilde{M}^-(d)$ when the order contribution is constant, while the largest value is obtained when only the inclusion of the largest order fully decreases the order contribution (i.e. it reduces $\tilde{M}^-(d)$ from $\tilde{M}_{max}$ to 0. This is the opposite of what discussed in the previous section.
By solving the right-hand side of the inequality, we obtain the largest value of the sum of the centered order contribution 

\begin{equation}
    0 < \sum_{d=d_{min}}^{d_{max}}\tilde{M}^-(d) \leq  \tilde{M}_{max}(d_{max} - d{_{min}})
\end{equation}
As we are not considering the monotonic case, we normalize each member of the inequality by $\tilde{M}_{max}$ and we obtain

\begin{equation}
    0 < \sum_{d=d_{min}}^{d_{max}}(\tilde{M}^-(d)/\tilde{M}_{max}) \leq  (d_{max} - d{_{min}})
\end{equation}
Finally, we normalize each member by $(d_{max} - d{_{min}-1})$ and obtain

\begin{equation}
    0 < \frac{\sum_{d=d_{min}}^{d_{max}}(\tilde{M}^-(d)/\tilde{M}_{max})}{d_{max} - d{_{min}}} \leq 1 
\end{equation}
We can thus define the order relevance for decreasing order contributions as 
\begin{equation}
 \Gamma^{dec}_H(M) =
  \begin{cases}
  0 & \text{for}\  \tilde{M} = 0 \\
  \frac{\sum_{d=d_{min}}^{d_{max}}(\tilde{M}^-(d)/\tilde{M}_{max})}{d_{max} - d{_{min}}} & \text{for}\ \tilde{M} \neq 0
  \end{cases}
\end{equation}
In this case, 1 is obtained when only the largest order contributes, while 0 is obtained when the order contribution is constant. This is the opposite of the interpretation of the order contribution for the monotonically increasing (or constant) order contribution.
Thus, large values of $\Gamma^{dec}_H(M)$ correspond to a small contribution of small order hyperlinks (and a corresponding large contribution of the large order ones), while small values indicate a large contribution of small orders (and a small contribution of large ones). 
Note that in the case of the constant order contribution, one can use in principle both order relevance formulas, obtaining the same conclusion. Indeed, if we have a constant order contribution, both formulas will suggest that the smallest size is the only one contributing to the considered network measure.

\subsection*{Effect of non-monotonicity on the measures of $\sigma_{\ell}$}
\label{sec:checks}
\begin{figure}
    \centering
    \includegraphics[width=0.7\linewidth]{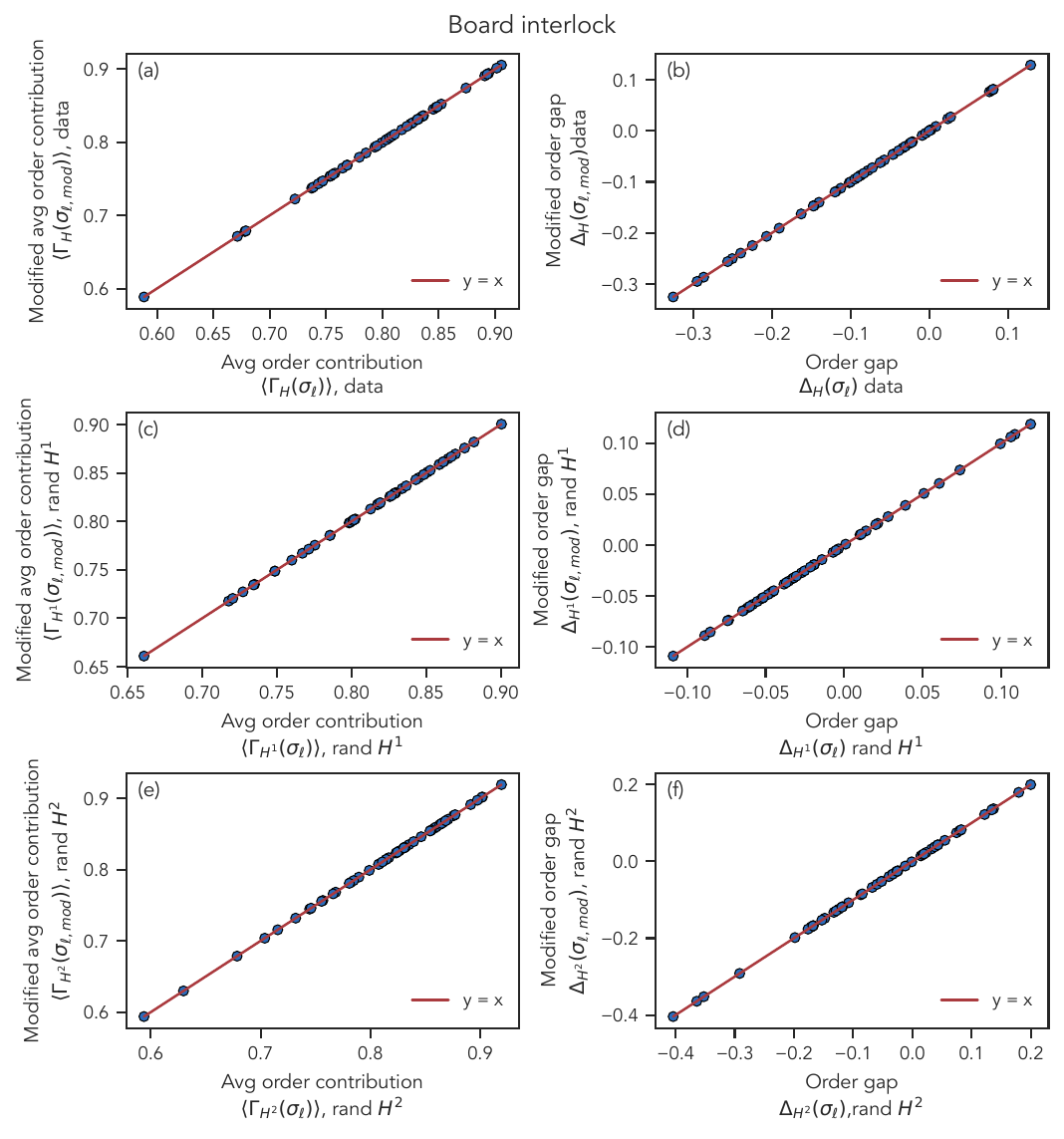}
   \caption{\textbf{Difference in the observed average order relevance $\langle \Gamma(\sigma_{\ell})\rangle$ (left) and in the order gap $\Delta(\sigma_{\ell})$} (right) obtained by following the original and modified method to compute the values of $\sigma^-{\ell}(d)$ and $\sigma^+{\ell}(d)$  \textbf{in the original board interlock network $H$ (a,b) and the two randomizations $H^1$} (c,d) \textbf{and $H^2$} (e,f).}
    \label{fig:board_check}
\end{figure}

\begin{figure}
    \centering
    \includegraphics[width=0.7\linewidth]{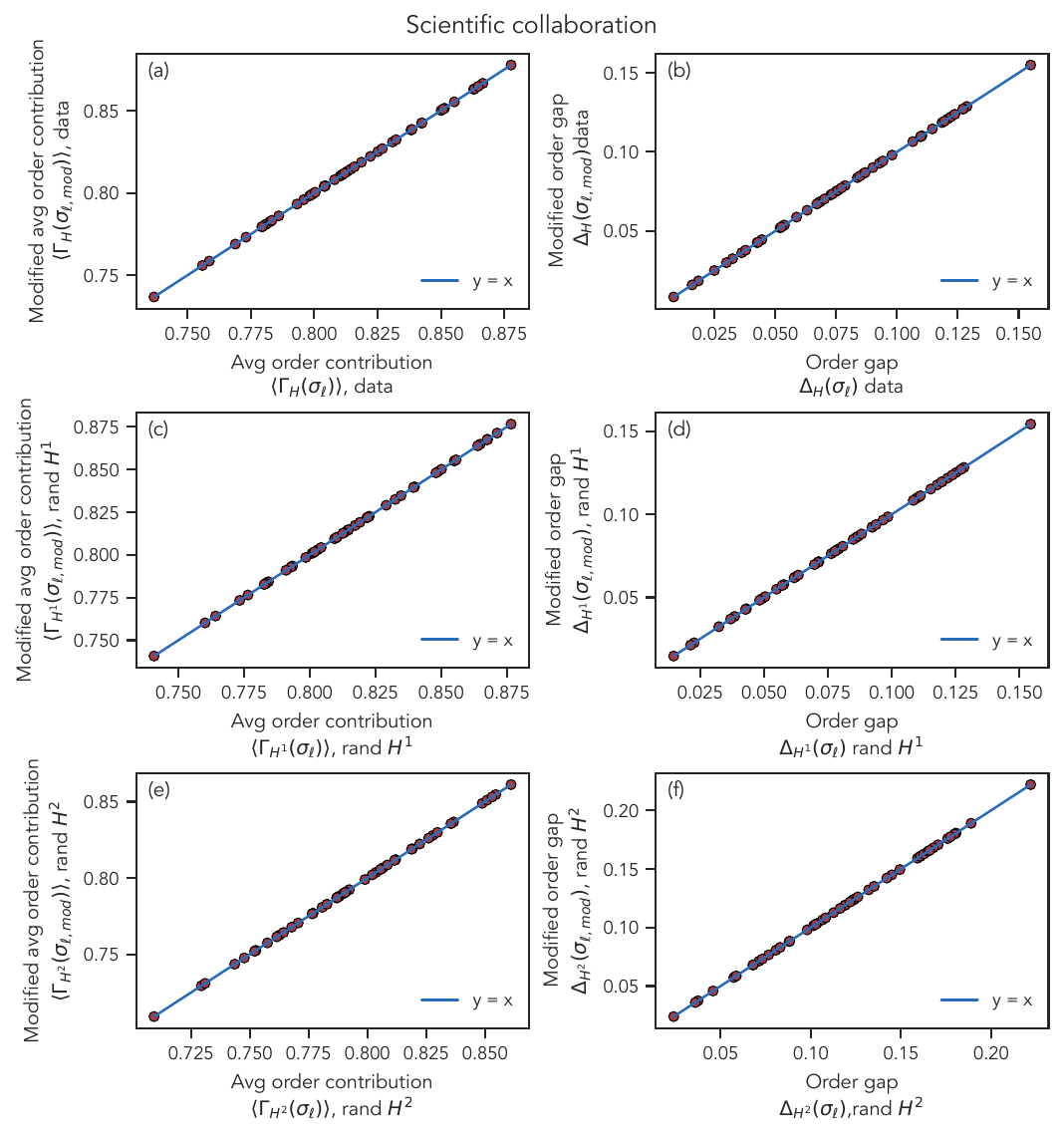}
   \caption{\textbf{Comparison of the values of the observed average order relevance $\langle \Gamma(\sigma_{\ell})\rangle$ (left) and in the order gap $\Delta(\sigma_{\ell})$} (right) obtained by following the original and modified method to compute the values of $\sigma^-{\ell}(d)$ and $\sigma^+{\ell}(d)$ \textbf{in the original scientific collaboration network $H$ (a,b) and the two randomizations $H^1$} (c,d) \textbf{and $H^2$} (e,f)}
    \label{fig:collaboration_check}
\end{figure}

As discussed in Section 2.4.3 of the main manuscript, the order contribution $\sigma_{\ell}^-(d)$ and the corresponding complementary $\overline{\sigma_{\ell}^-(d)}$ could not be monotonic functions of the order $d$. Each $\sigma_{\ell}^-(d)$ and $\sigma_{\ell}^+(d)$ are the number of nodes with label $\ell$ in the connected component of $H^-(d)$ and $H^+(d)$, respectively. A sufficient condition for each couple of orders $d \geq d'$ to have $\sigma_{\ell}^-(d)\geq \sigma_{\ell}^-(d')$ is that the set of nodes in the largest connected component of $H^-(d)$ includes all nodes in the largest connected component of $H^-(d')$ (condition (a)).
Equivalently, for each couple of orders $d \leq d'$,  $\sigma_{\ell}^+(d)\geq \sigma_{\ell}^+(d')$ holds if the largest connected component of $H^+(d)$ includes all nodes of the one of $H^+(d')$ (condition (b)). Thus, as long as these two conditions hold for each pair of orders $d,d'$, the corresponding $\sigma^-_{\ell}(d)$ and $\sigma^+_{\ell}(d)$ in the network are monotonic.
We checked if these conditions hold for each real network $H$ and the corresponding randomized null models $H^1$ and $H^2$ for both board interlock and scientific collaboration
network. 
We discovered that for the scientific collaboration network, both conditions (a) and (b) hold for $H$, $H^1$, and $H^2$.
This is, however not the case for the board interlock network. In this case, condition (a) is always satisfied, while condition (b) is not satisfied at large orders in $H$ and $H^2$.
This produces a non-monotonic order contribution and inverse order contribution.
To quantify how this non-monotonicity affects our results, we compared the values of the order gap and the average order relevance at each country with the results obtained by enforcing both conditions (a) and (b) in each network.
To enforce these conditions of inclusion, we computed the values $\sigma^-_{\ell,mod}(d)$ and $\sigma^+_{\ell,mod}(d)$ with the following method.
For $\sigma^-_{\ell,mod}(d)$, we started by computing the largest connected component of $H^-(d_{max})$. Note that $\sigma^-_{\ell,mod}(d_{max}) = \sigma^-_{\ell}(d_{max})$ by definition. Then, for $d_{max}-1$, we defined $\sigma^-_{\ell,mod}(d_{max}-1)$ as the size of the connected component with the maximum number of nodes also present in $\sigma^-_{\ell,mod}(d_{max})$. By repeating this procedure for each couple of orders $d,d'$ until $d_{min}+1,d_{min}$, we enforce condition (a).
Similarly, for $\sigma^+_{\ell,mod}(d)$, we started by computing the largest connected component of $H^+(d_{min})$. Also here, $\sigma^+_{\ell,mod}(d_{min}) = \sigma^+_{\ell}(d_{min})$ by definition. Then, for $d_{min}+1$, we defined $\sigma^+_{\ell,mod}(d_{min}+1)$ as the size of the connected component with the maximum number of nodes also present in $\sigma^+_{\ell,mod}(d_{min})$. By repeating this procedure for each couple of orders $d,d'$ until $d_{max}-1,d_{max}$, we enforce condition (b).
Finally, from the corrected order contribution $\sigma^-_{\ell,mod}(d)$ and the inverse $\sigma^+_{\ell,mod}(d)$, we follow the usual procedure to compute the corrected corresponding average order relevance and order gap.
To support our claim that no substantial difference in results is obtained by computing the order contributions with the two different methods, in Figures~\ref{fig:board_check} and~\ref{fig:collaboration_check}, we compared the average order relevance $\langle \Gamma (\sigma_{\ell,mod})\rangle$ and the order gap $\Delta(\sigma_{\ell,mod})$ with their original values $\langle \Gamma (\sigma_{\ell})\rangle$ and $\Delta(\sigma_{\ell})$.

\newpage

\subsection*{Effect of randomizations on the group balance}

\begin{figure}[h!]
    \centering
    \includegraphics[width=0.8\linewidth]{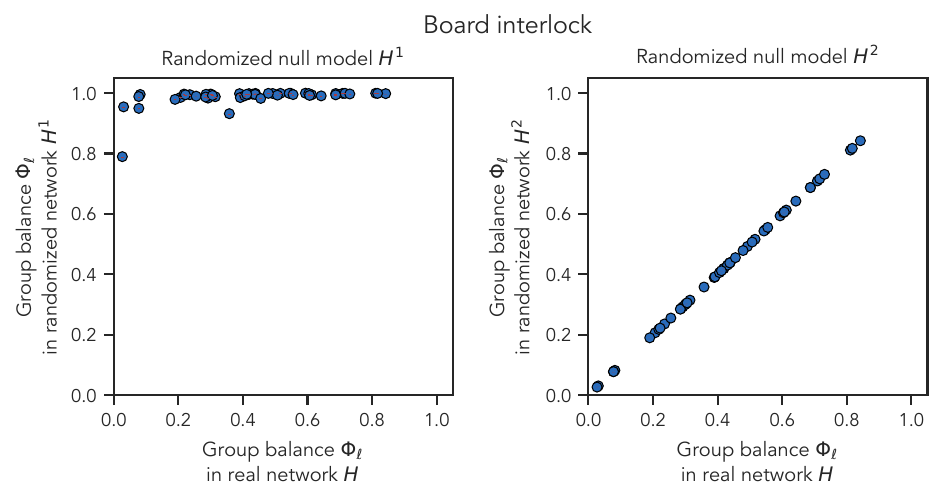}
    \caption{\textbf{Randomized group balance vs original group balance for $H^1$ (a) and $H^2$ (b)} of the 50 considered countries in the board interlock network. Each value is reported is an average with the corresponding standard deviation obtained from 10 independent realizations of the considered randomized model.}
    \label{fig:local_component_board_rand2}
\end{figure}

\begin{figure}[h!]
    \centering
    \includegraphics[width=0.8\linewidth]{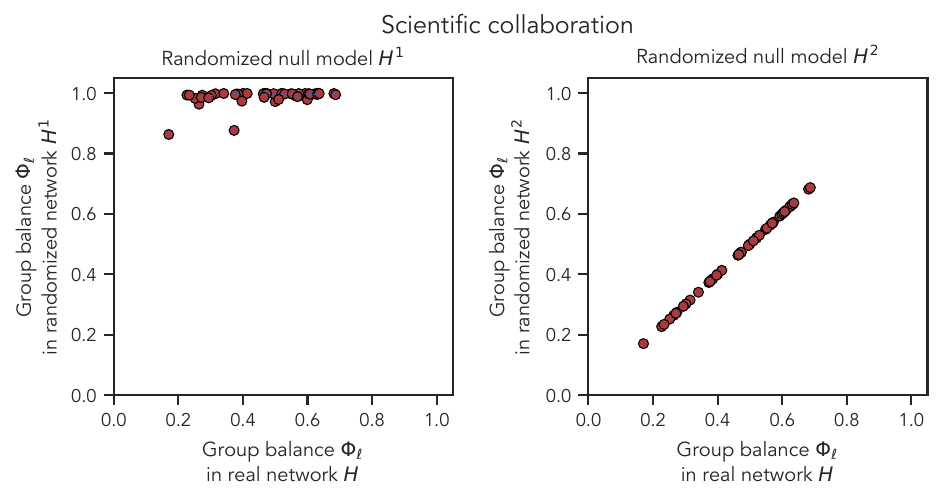}
    \caption{\textbf{Randomized group balance vs original group balance for $H^1$ (a) and $H^2$ (b)} of the 50 considered countries in the scientific collaboration network. Each value is reported is an average with the corresponding standard deviation obtained from 10 independent realizations of the considered randomized model.}
    \label{fig:local_component_board_rand2}
\end{figure}
\newpage
\subsection*{World map of the order relevance of the node labels' composition in the largest connected component}

\begin{figure}[h!]
    \centering
    \includegraphics[width=0.54\linewidth]{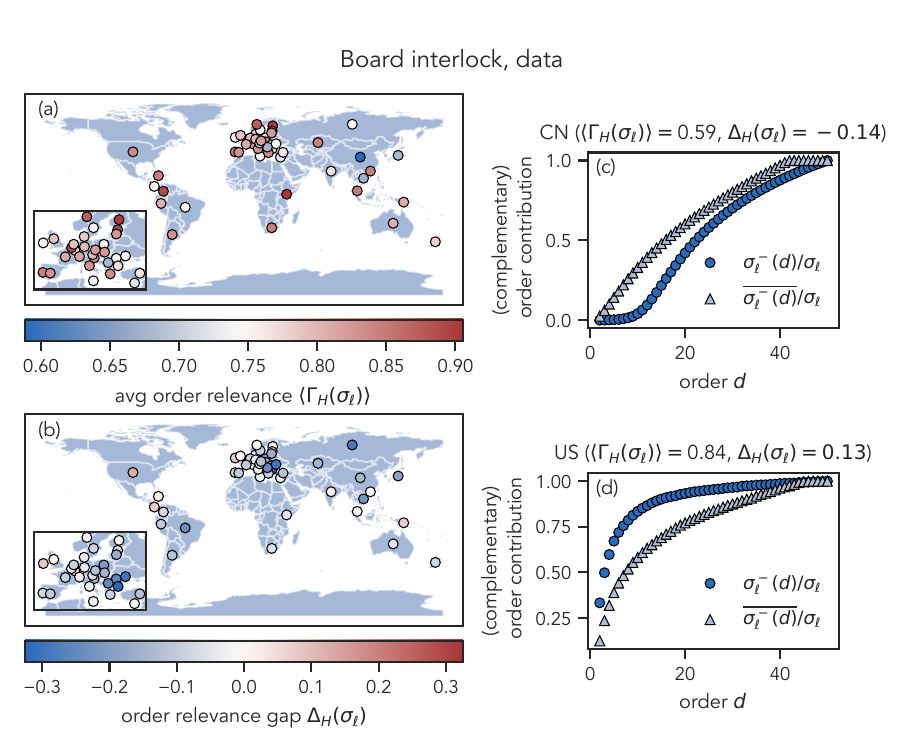}
    \caption{\textbf{Average order relevance (a) and order gap (c) obtained from the board interlock network} of the normalized number of nodes with a given label in the largest connected component $\Delta(\sigma_{\ell})$ for each country, together with the order contribution $\sigma^-_{\ell}(d)$ (dark blue) and the corresponding complementary $\overline{\sigma^-_{\ell}(d)}$ (light blue) of China (b) and United States (d).}
    \label{fig:local_component_board_rand}
\end{figure}

\begin{figure}[h!]
    \centering
    \includegraphics[width=0.54\linewidth]{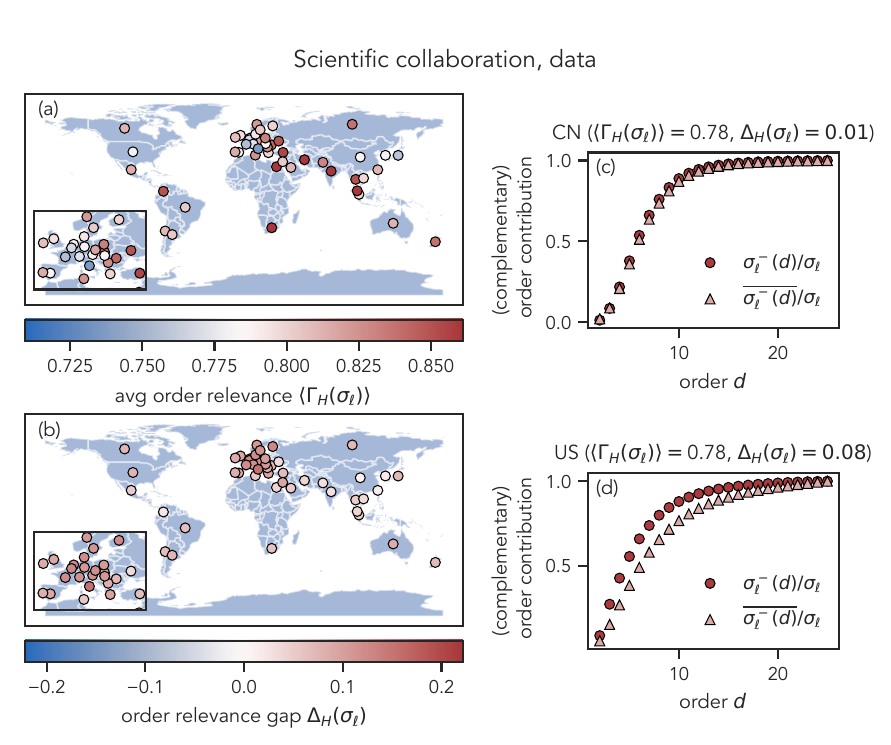}
     \caption{\textbf{Average order relevance (a) and order gap (c) obtained from the scientific collaboration network} of the normalized number of nodes with a given label in the largest connected component $\Delta(\sigma_{\ell})$ for each country, together with the order contribution  $\sigma^-_{\ell}(d)$ (dark red) and the corresponding complementary $\overline{\sigma^-_{\ell}(d)}$ (light red) of China (b) and United States (d). }
    \label{fig:local_component_board_rand1}
\end{figure}
\newpage
\subsection*{Effect of randomizations on the node labels’ composition in the largest connected component}

\begin{figure}[h!]
    \centering
    \includegraphics[width=0.54\linewidth]{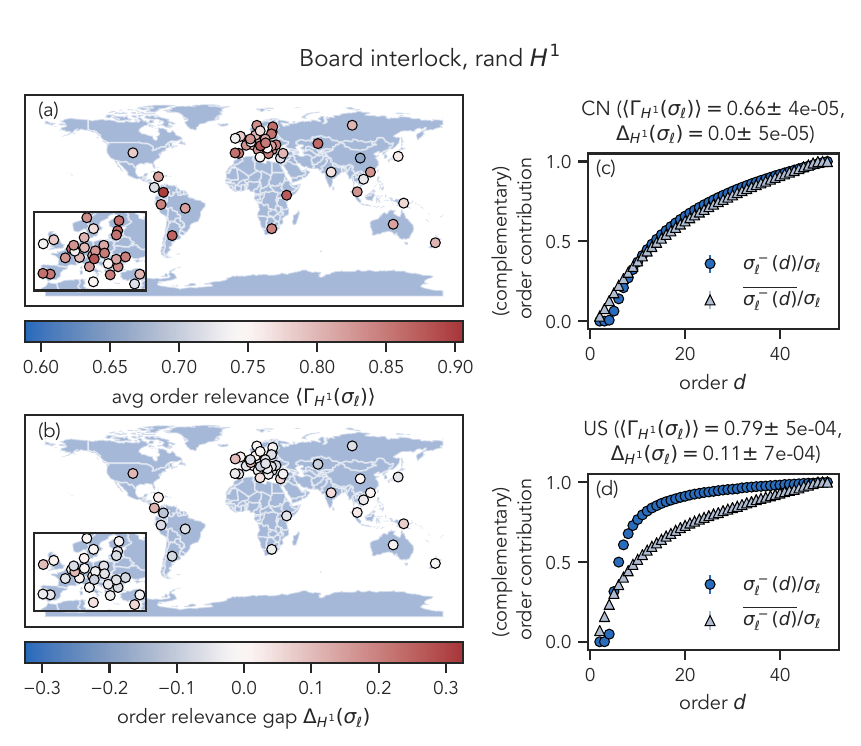}
    \caption{\textbf{Average order relevance (a) and order gap (c) in the randomized null model $H^1$ obtained from the board interlock network} of the normalized number of nodes with a given label in the largest connected component $\Delta(\sigma_{\ell})$ for each country, together with the order contribution $\sigma^-_{\ell}(d)$ and the corresponding complementary $\overline{\sigma^-_{\ell}(d)}$ of China (b) and United States (d). Each value is reported is an average with the corresponding standard deviation obtained from 10 independent realizations of the considered randomized model.}
    \label{fig:local_component_board_rand2}
\end{figure}

\begin{figure}[h!]
    \centering
    \includegraphics[width=0.54\linewidth]{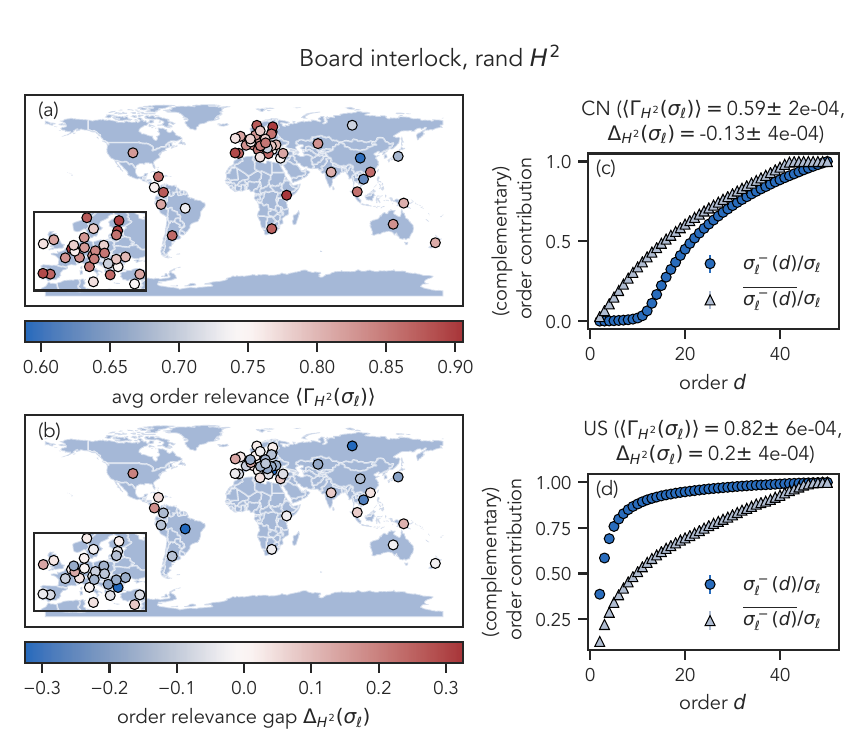}
     \caption{\textbf{Average order relevance (a) and order gap (c) in the randomized null model $H^2$ obtained from the board interlock network} of the normalized number of nodes with a given label in the largest connected component $\Delta(\sigma_{\ell})$ for each country, together with the order contribution $\sigma^-_{\ell}(d)$ and the corresponding complementary $\overline{\sigma^-_{\ell}(d)}$ of China (b) and United States (d). Each value is reported is an average with the corresponding standard deviation obtained from 10 independent realizations of the considered randomized model.}
    \label{fig:local_component_board_rand1}
\end{figure}

\begin{figure}[h!]
    \centering
    \includegraphics[width=0.54\linewidth]{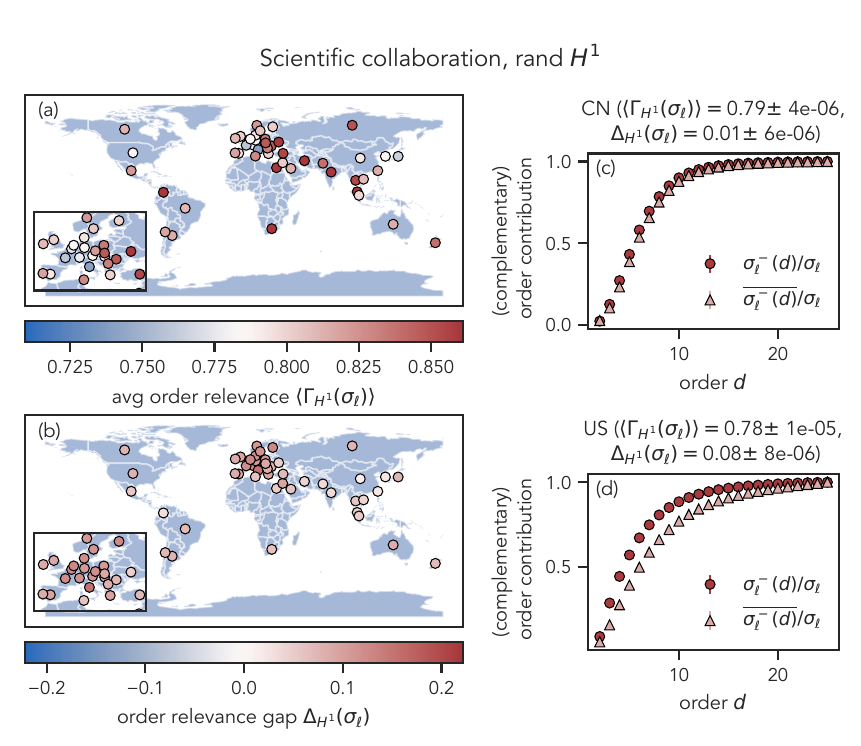}
    \caption{\textbf{Average order relevance (a) and order gap (c) in the randomized null model $H^1$ obtained from the scientific collaboration network} of the normalized number of nodes with a given label in the largest connected component $\Delta(\sigma_{\ell})$ for each country, together with the order contribution $\sigma^-_{\ell}(d)$ and the corresponding complementary $\overline{\sigma^-_{\ell}(d)}$ of China (b) and United States (d). Each value is reported is an average with the corresponding standard deviation obtained from 10 independent realizations of the considered randomized model.}
    \label{fig:local_component_collaboration_rand2}
\end{figure}

\begin{figure}[h!]
    \centering
    \includegraphics[width=0.54\linewidth]{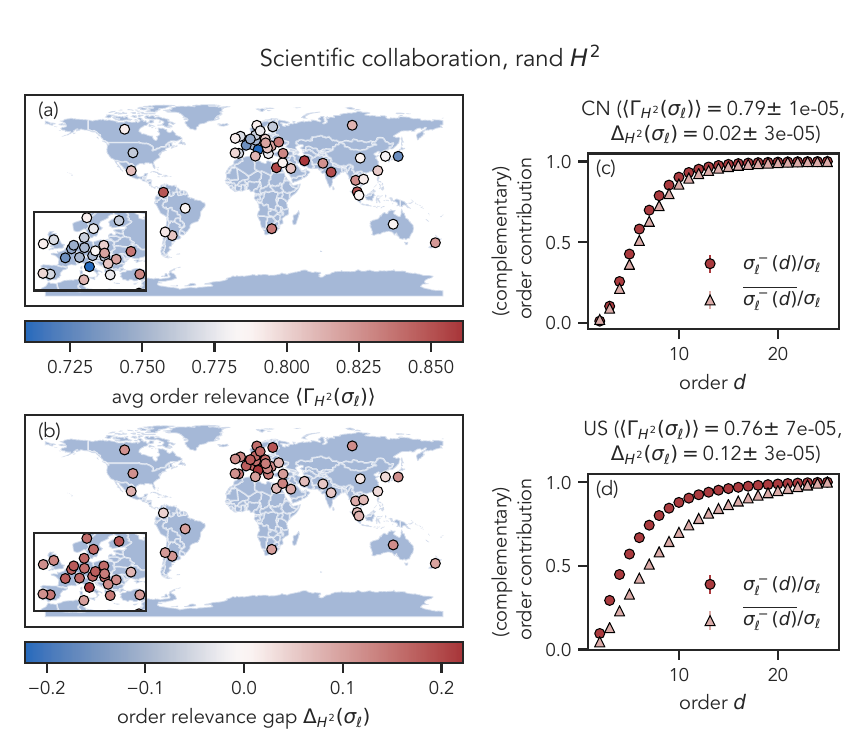}
   \caption{\textbf{Average order relevance (a) and order gap (c) in the randomized null model $H^2$ obtained from the scientific collaboration network} of the normalized number of nodes with a given label in the largest connected component $\Delta(\sigma_{\ell})$ for each country, together with the order contribution $\sigma^-_{\ell}(d)$ and the corresponding complementary $\overline{\sigma^-_{\ell}(d)}$ of China (b) and United States (d). Each value is reported is an average with the corresponding standard deviation obtained from 10 independent realizations of the considered randomized model.}
    \label{fig:local_component_collaboration_rand1}
\end{figure}

\clearpage 
\subsection*{Impact of randomization, scatter plot}
\begin{figure}[h!]
    \centering
    \includegraphics[width=0.55\linewidth]{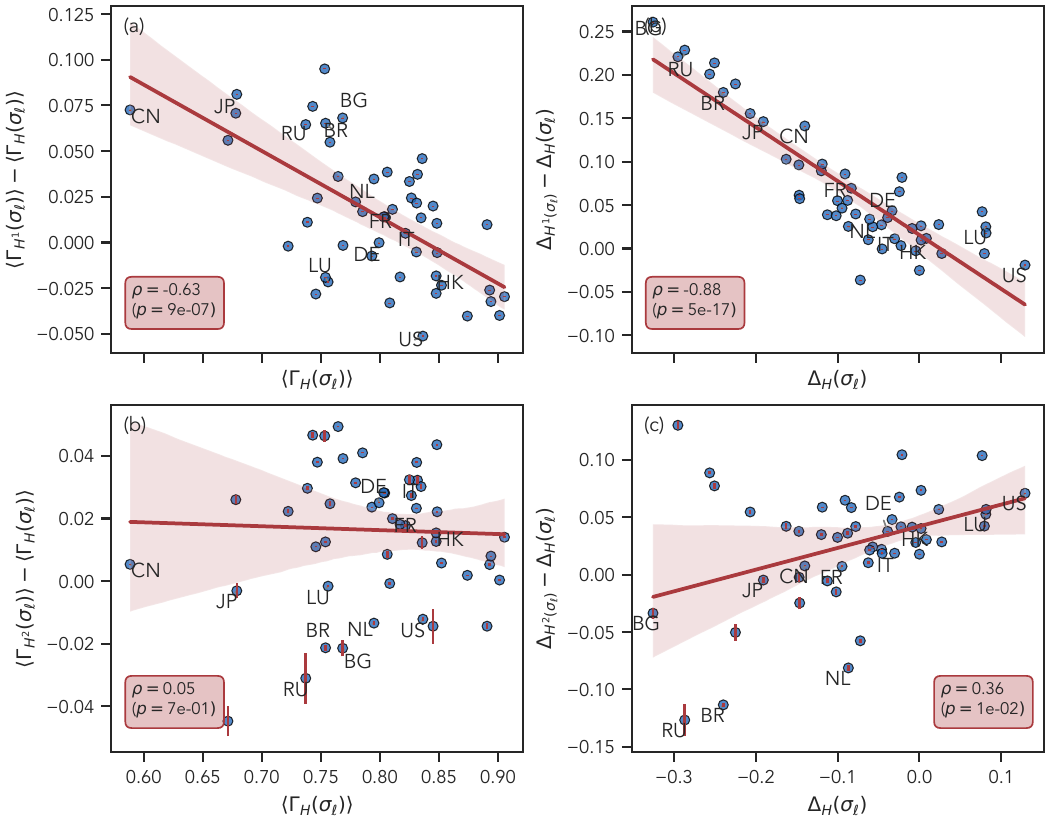}
   \caption{\textbf{Difference in the values of average order relevance $\langle \Gamma(\sigma_{\ell})\rangle$} (a,c) \textbf{and in the order gap $\Delta(\sigma_{\ell})$} (b,d) of the number of nodes with a given label in the largest connected component $\sigma_{\ell}$ \textbf{computed from the original board interlock network $H$ and the two randomizations $H^1$} (a,b) \textbf{and $H^2$} (c,d). Each value is reported is an average with the corresponding standard deviation obtained from 10 independent realizations of the considered randomized model. We further show the linear fit and its corresponding 95\% confidence intervals, together with the value of the linear correlation coefficient $\rho$ and corresponding $p$-value.}
    \label{fig:rand_comparison_board}
\end{figure}

\begin{figure}[h!]
    \centering
    \includegraphics[width=0.55\linewidth]{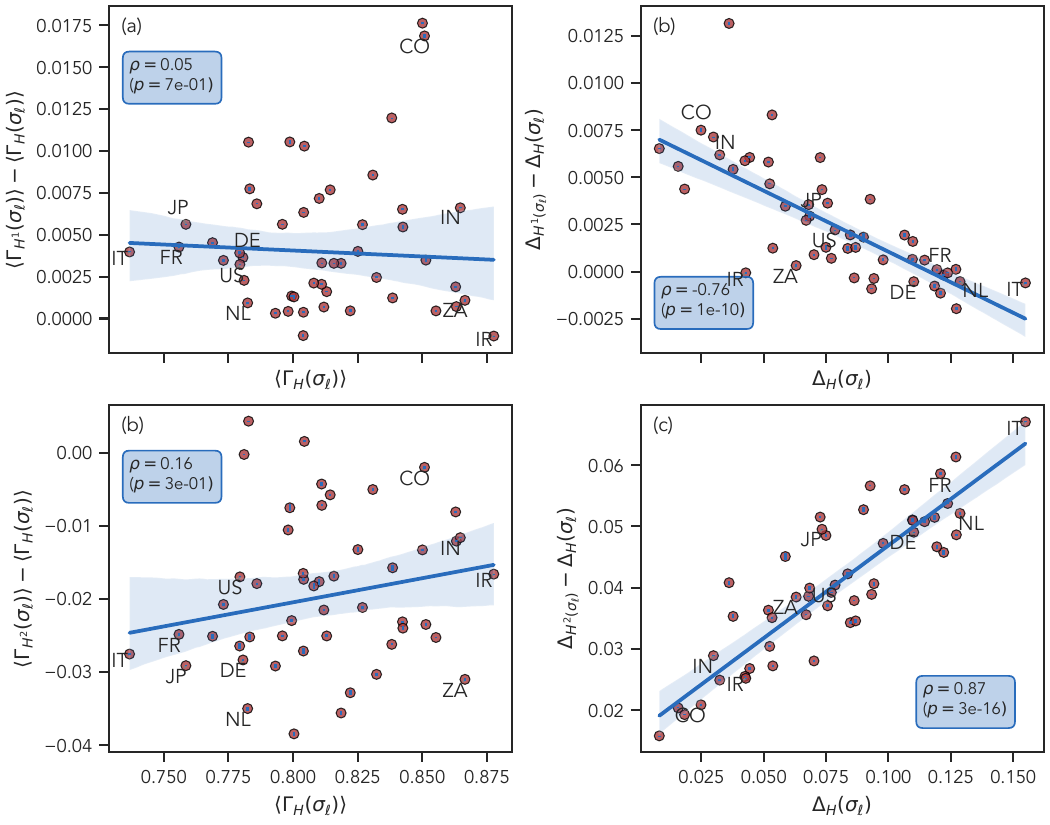}
   \caption{\textbf{Difference in the values of average order relevance $\langle \Gamma(\sigma_{\ell})\rangle$} (a,c) \textbf{and in the order gap $\Delta(\sigma_{\ell})$} (b,d) of the number of nodes with a given label in the largest connected component $\sigma_{\ell}$ \textbf{computed from the original scientific collaboration network $H$ and the two randomizations $H^1$} (a,b) \textbf{and $H^2$} (c,d). Each value is reported is an average with the corresponding standard deviation obtained from 10 independent realizations of the considered randomized model. We further show the linear fit and its corresponding 95\% confidence intervals, together with the value of the linear correlation coefficient $\rho$ and corresponding $p$-value.}
    \label{fig:check_board}
\end{figure}

\clearpage 

\subsection*{Correlation analysis of the group balance and order relevance gap against aggregated national metrics}
\begin{figure}[h]
    \centering
    \includegraphics[width=0.97\linewidth]{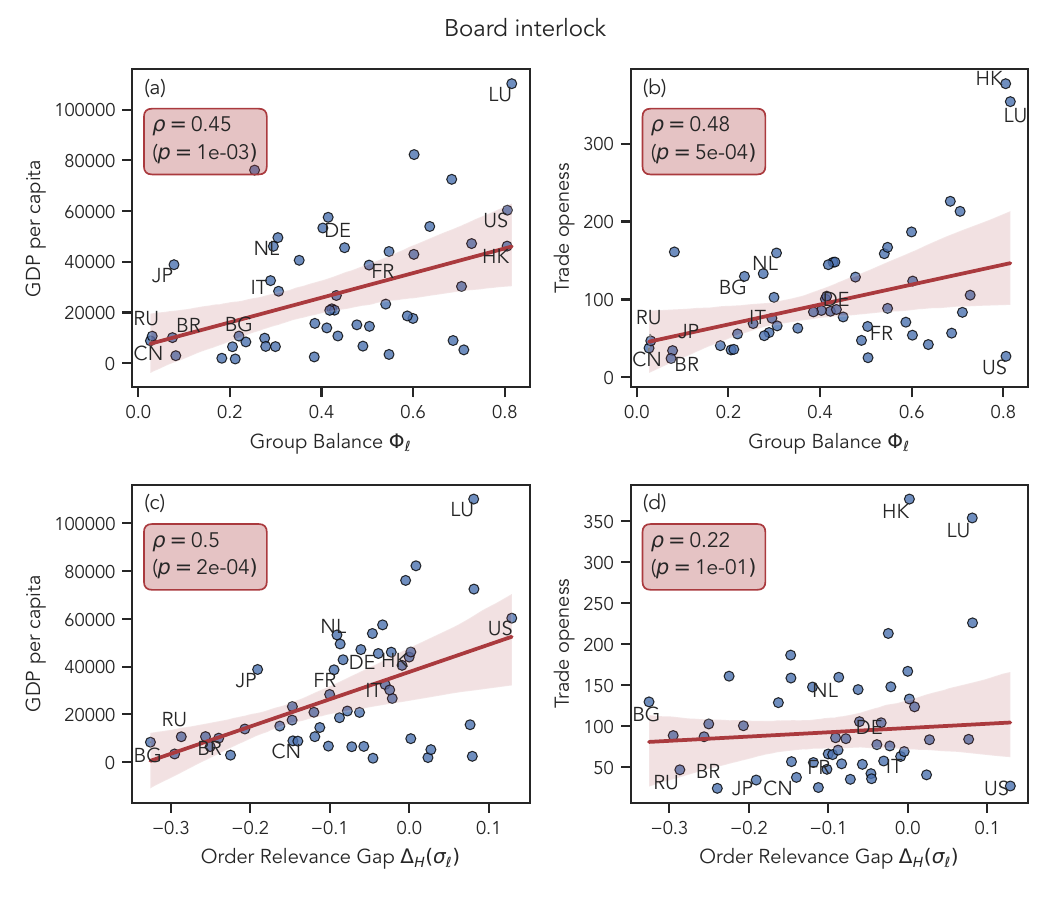}
   \caption{\textbf{Correlation analysis of the group balance} and the corresponding \textbf{GDP per capita} (a) and \textbf{trade openness} (b) of the 50 considered countries of the \textbf{board interlock network}.We further show the linear fit and its corresponding 95\% confidence intervals, together with the value of the linear correlation coefficient $\rho$ and corresponding $p$-value.}
    \label{fig:correlation_group_board}
\end{figure}

In Sections 4.2.2 and 4.2.3 of the main manuscript we have discussed the results obtained from the analysis of the group balance and the corresponding order relevance gap of the number of nodes $\sigma_{\ell}$ of a given country $\ell$ connected to the largest connected component. In this section, we investigate whether large or small values of group balance and order relevance gap are associated with specific economic or academic characteristics of the considered countries. We focus in particular on three different metrics: 
\begin{itemize}
    \item GDP per capita: it is the amount of Gross Domestic Product normalized by the number of inhabitants of a country. We use this as a proxy of the overall wealth of a given country
    \item Trade openness: it is a measure of the orientation of a country to trade internationally
    \item Academic freedom index: it measures the level of academic freedom of a country based on five indicators: freedom to research and teach; freedom of academic exchange and dissemination; institutional autonomy; campus integrity; and freedom of academic and cultural expression.
\end{itemize}
Data about GDP per capita and trade openness were obtained from the World Bank Open Data website \url{https://data.worldbank.org/}, while the Academic freedom index was obtained from the V-Dem dataset \url{https://v-dem.net/data/the-v-dem-dataset/}, the world's most comprehensive and detailed dataset of democracy ratings.
We used the values obtained from 2017, which is the last year included in the ORBIS dataset, which was used to construct the board interlock network. 
Figures~\ref{fig:correlation_group_board}a-b show that, in the board interlock network, the group balance correlates with both the GDP per capita and trade openness. This suggests that, in general, the countries that are either richer or more oriented to international trade are those in which the corresponding companies tend to be more often interlocked with other international companies by directors. 

\begin{figure}[!h]
    \centering
    \includegraphics[width=0.97\linewidth]{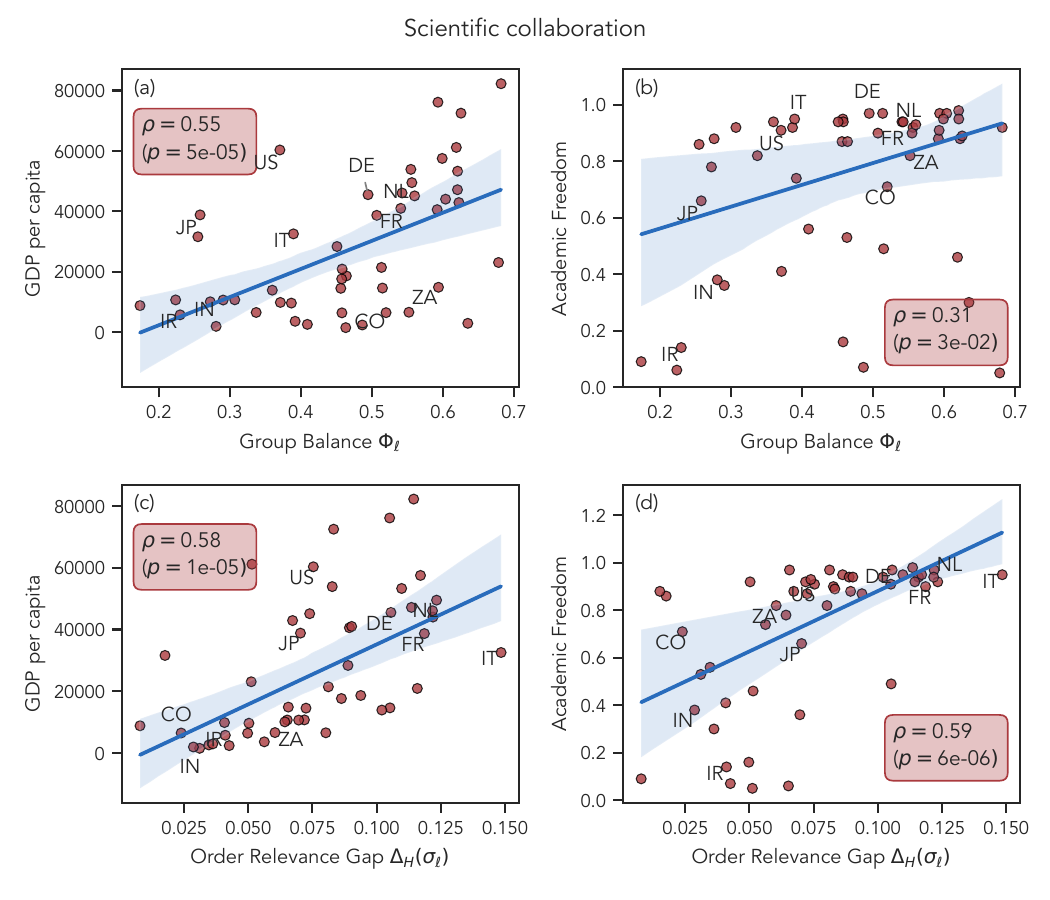}
   \caption{\textbf{Correlation analysis of the order relevance gap and $\Delta_H(\sigma_{\ell})$} and of the number of companies based in a given country belonging to the largest connected component and the corresponding \textbf{GDP per capita} (a) and \textbf{trade openness} (b) of the 50 considered countries of the \textbf{board interlock network}. We further show the linear fit and its corresponding 95\% confidence intervals, together with the value of the linear correlation coefficient $\rho$ and corresponding $p$-value.}
    \label{fig:correlation_group_collaboration}
\end{figure}

In Figures~\ref{fig:correlation_group_board}c-d, we observe that the order relevance gap is positively correlated to the GDP per capita, but not with the trade openness ($p$-value > 0.05). This suggests that the rich countries tend to also be integrated in the overall largest connected component of the board interlock more redundantly. However, this does not hold for countries with a larger tendency to international trade.
We further observe that, in the scientific collaboration network, the  countries with a high GDP per capita tend to have larger group balance and group balance, as shown by the positive correlation observed in Figures~\ref{fig:correlation_group_collaboration}a-c, suggesting that researchers based in these countries tend to be integrated more redundantly to the overall largest connected component and are more likely engaged in international collaborations. A larger redundancy is also observed in countries with a higher academic freedom 
(Figure~\ref{fig:correlation_group_collaboration}d), as shown by the positive correlation between the order relevance gap and the academic freedom index. Differently, only a weak positive correlation is observed between the group balance of a country and its academic freedom.
In both datasets, we observe that richer countries tend to have a high group balance and are integrated in the largest connected component in a redundant way, while countries with a lower GDP per capita seems to have a lower group balance and are integrated to the network in a more synergistic way. 
Such results overall show that the differences in the group balance and order relevance gap, partially correspond to specific overall economic and academic differences of the considered coutries.

